\PassOptionsToPackage{dvipsnames}{xcolor}
\documentclass[a4paper,twocolumn,superscriptaddress,preprintnumbers,nobalancelastpage,longbibliography,accepted=2022-08-31]{quantumarticle}

 \pdfoutput=1
\usepackage{graphicx, xcolor, graphpap}
\usepackage{enumitem}
\usepackage{amssymb}
\usepackage{amsthm}
\usepackage{multirow}
\usepackage[colorlinks=true, citecolor=RawSienna, linkcolor=MidnightBlue, urlcolor=MidnightBlue, linktocpage=true]{hyperref}

\usepackage[numbers]{natbib}

\usepackage[T1]{fontenc}

\usepackage{bbm}
\usepackage{thmtools,thm-restate}
\usepackage{verbatim}
\usepackage{mathtools}
\usepackage{titlesec}
\usepackage{amsmath}
\usepackage[linesnumbered,ruled,vlined]{algorithm2e}
\SetKwInput{kwInit}{Init}

\usepackage{tikz}
\usetikzlibrary{shapes.geometric, arrows}
\tikzstyle{startstop} = [rectangle, rounded corners, minimum width=3cm, minimum height=1cm,text centered, draw=black, fill=red!30]
\tikzstyle{io} = [trapezium, trapezium left angle=70, trapezium right angle=110, minimum width=3cm, minimum height=1cm, text centered, draw=black, fill=blue!30]
\tikzstyle{process} = [rectangle, minimum width=3cm, minimum height=1cm, text centered, draw=black, fill=orange!30]
\tikzstyle{decision} = [diamond, minimum width=3cm, minimum height=1cm, text centered, draw=black, fill=green!30]
\tikzstyle{arrow} = [thick,->,>=stealth]

\long\def\ca#1\cb{} 

\newcommand{\ket}[1]{|#1\rangle}               
\newcommand{\bra}[1]{\langle #1|}              
\newcommand{\dya}[1]{\ket{#1}\!\bra{#1}}

\newcommand{\opt}{\text{opt}}

\newcommand{\BC}{\mathcal{B}}
\newcommand{\CC}{\mathcal{C}}

\newcommand{\GC}{\mathcal{G}}

\newcommand{\NC}{\mathcal{N}}
\newcommand{\OC}{\mathcal{O}}
\newcommand{\PC}{\mathcal{P}}

\newcommand{\SC}{\mathcal{S}}

\newcommand{\UC}{\mathcal{U}}
\newcommand{\VC}{\mathcal{V}}

\newcommand{\Tr}{{\rm Tr}}

\renewcommand{\leq}{\leqslant}

\renewcommand{\vec}[1]{\boldsymbol{#1}}  

\newcommand{\ad}{^\dagger}

\newcommand*{\id}{\openone}

\newcommand{\thv}{\vec{\theta}}
\newcommand{\gamv}{\vec{\gamma}}

{}
{}
\newtheorem{theorem}{Theorem}

\newtheorem{corollary}{Corollary}
\newtheorem{proposition}{Proposition}

\newtheorem{definition}{Definition}

\begin{document}

\title{Non-trivial symmetries in quantum landscapes and their resilience to quantum noise}

\author{Enrico Fontana}
\affiliation{Theoretical Division, Los Alamos National Laboratory, Los Alamos, NM 87545, USA}
\affiliation{Department of Computer and Information Sciences, University of Strathclyde, 26 Richmond Street, Glasgow  G1 1XH, UK}
\affiliation{National Physical Laboratory, Teddington  TW11 0LW, UK}

\author{M. Cerezo}
\affiliation{Theoretical Division, Los Alamos National Laboratory, Los Alamos, NM 87545, USA}
\affiliation{Center for Nonlinear Studies, Los Alamos National Laboratory, Los Alamos, NM, USA
}

\author{Andrew Arrasmith}
\affiliation{Theoretical Division, Los Alamos National Laboratory, Los Alamos, NM 87545, USA}

\author{Ivan Rungger}
\affiliation{National Physical Laboratory, Teddington, UK}

\author{Patrick J. Coles}
\affiliation{Theoretical Division, Los Alamos National Laboratory, Los Alamos, NM 87545, USA}

\begin{abstract}
Very little is known about the cost landscape for Parametrized Quantum Circuits (PQCs). Nevertheless, PQCs are employed in Quantum Neural Networks and Variational Quantum Algorithms, which may allow for near-term quantum advantage. Such applications require good optimizers to train PQCs. Recent works have focused on quantum-aware optimizers specifically tailored for PQCs. However, ignorance of the cost landscape could hinder progress towards such optimizers. 
In this work, we analytically prove two results for PQCs: (1) We find an exponentially large symmetry in PQCs, yielding an exponentially large degeneracy of the minima in the cost landscape. Alternatively, this can be cast as an exponential reduction in the volume of relevant hyperparameter space. (2) We study the resilience of the symmetries under noise, and show that while it is conserved under unital noise, non-unital channels can break these symmetries and lift the degeneracy of minima, leading to multiple new local minima. Based on these results, we introduce an optimization method called Symmetry-based Minima Hopping (SYMH), which exploits the underlying symmetries in PQCs. Our numerical simulations show that SYMH improves the overall optimizer performance in the presence of non-unital noise at a level comparable to current hardware.
Overall, this work derives large-scale circuit symmetries from local gate transformations, and uses them to construct a noise-aware optimization method.

\end{abstract}

\maketitle

\section{Introduction}

The era of Noisy Intermediate Scale Quantum (NISQ)\cite{preskill2018quantum} computing has led to the emergence of novel algorithmic paradigms. Arguably, the leading role has been played by Parametrized Quantum Circuits (PQCs), which are exploited for both Variational Quantum Algorithms~\cite{cerezo2020variationalreview,VQE,mcclean2016theory,qaoa2014,Romero,Khatri2019quantumassisted,VQSD,arrasmith2019variational,cerezo2020variationalfidelity,cirstoiu2020variational,bravo-prieto2019,cerezo2020variational,rungger2019dynamical} and Quantum Neural Networks~\cite{schuld2014quest,cong2019quantum,beer2020training,verdon2018universal,patterson2019quantum}. Training PQCs involves a hybrid quantum-classical optimization loop. Typically, the problem is encoded in a cost (or loss) function that is ideally efficient to evaluate on a quantum computer, but computationally expensive for a classical one. In practice, the cost function is estimated via measurements on a quantum computer which are usually post-processed on a classical device. While the quantum computer is used to evaluate the cost, a classical optimizer updates some (usually continuous) parameters associated with the quantum operations. PQCs with fixed gate structure are often referred to as a variational ansatz. 

High performance classical optimizers are crucial to successfully train PQCs. To aid in optimizer selection and development, a fair amount of work has gone into determining the nature of quantum variational cost landscapes~\cite{Huembeli2020Characterizing}. Important contributions include the development of analytical expressions for gradients of all orders~\cite{mitarai2018quantum,schuld2019evaluating,mitarai2019methodology,cerezo2020impact,mari2020estimating}, bounds on those derivatives~\cite{kubler2019adaptive}, and even explicit functional forms for the expectation values computed with PQCs~\cite{nakanishi2020sequential}. In addition, some light has recently been shed on the scaling of the gradient of quantum cost functions through a result known as barren plateaus \cite{mcclean2018barren,cerezo2020cost,sharma2020trainability,cerezo2020impact,holmes2020barren,pesah2020absence,marrero2020entanglement}. This demonstrates that the landscape flattens exponentially with problem size for deep, unstructured PQCs, and also for shallow PQCs with global cost functions.

Even if one manages to avoid these barren plateaus, there is an additional difficulty of optimizing in the presence of the hardware noise that defines NISQ devices~\cite{hamilton2020scalable}. Hardware noise is expected to modify the cost landscape, and indeed it was recently shown to produce a novel kind of barren plateau whose impact increases with PQC depth~\cite{wang2020noise}. Furthermore, while some models of hardware noise have been shown to leave the optimal parameters unchanged~\cite{sharma2019noise}, this does not hold in general~\cite{fontana2020evaluating}.

These results collectively reveal that the optimization of noisy PQCs presents many novel and unexpected challenges that must be addressed. This has spawned the field of quantum-aware optimizers, where researchers are developing classical optimizers that are specifically tailored to the unusual landscape issues in the quantum setting. Examples include quantum natural gradient~\cite{stokes2020quantum,koczor2019quantum}, sequential function fitting~\cite{nakanishi2019}, shot-frugal stochastic gradient descent~\cite{kubler2019adaptive,arrasmith2020operator,sweke2020stochastic}, landscape modeling~\cite{sung2020using}, and others~\cite{lavrijsen2020classical}. Many of these optimizers are good at finding a local minimum, which has been shown to be accelerated by gradient information~\cite{harrow2019}. However, there remains the question of how to escape or move between such minima to find the global minimum.

In this work, we present a technique for training PQCs that we call Symmetry-based Minima Hopping (SYMH, pronounced ``Sim''). As the name suggests, SYMH is a method for hoping between local minima that exploits underlying symmetries in PQCs. The SYMH technique can be combined with other optimizers, such as those in Refs.~\cite{stokes2020quantum,koczor2019quantum,nakanishi2019,kubler2019adaptive,arrasmith2020operator,sweke2020stochastic,sung2020using,lavrijsen2020classical}, to construct optimizers that search for minima that achieve lower costs by lowering the impact of noise. In this sense, our method is complementary to previous work, as classical local optimizers are easily integrated into the SYMH framework.

\begin{figure}[t]
    \centering
    \includegraphics[width=.95\columnwidth]{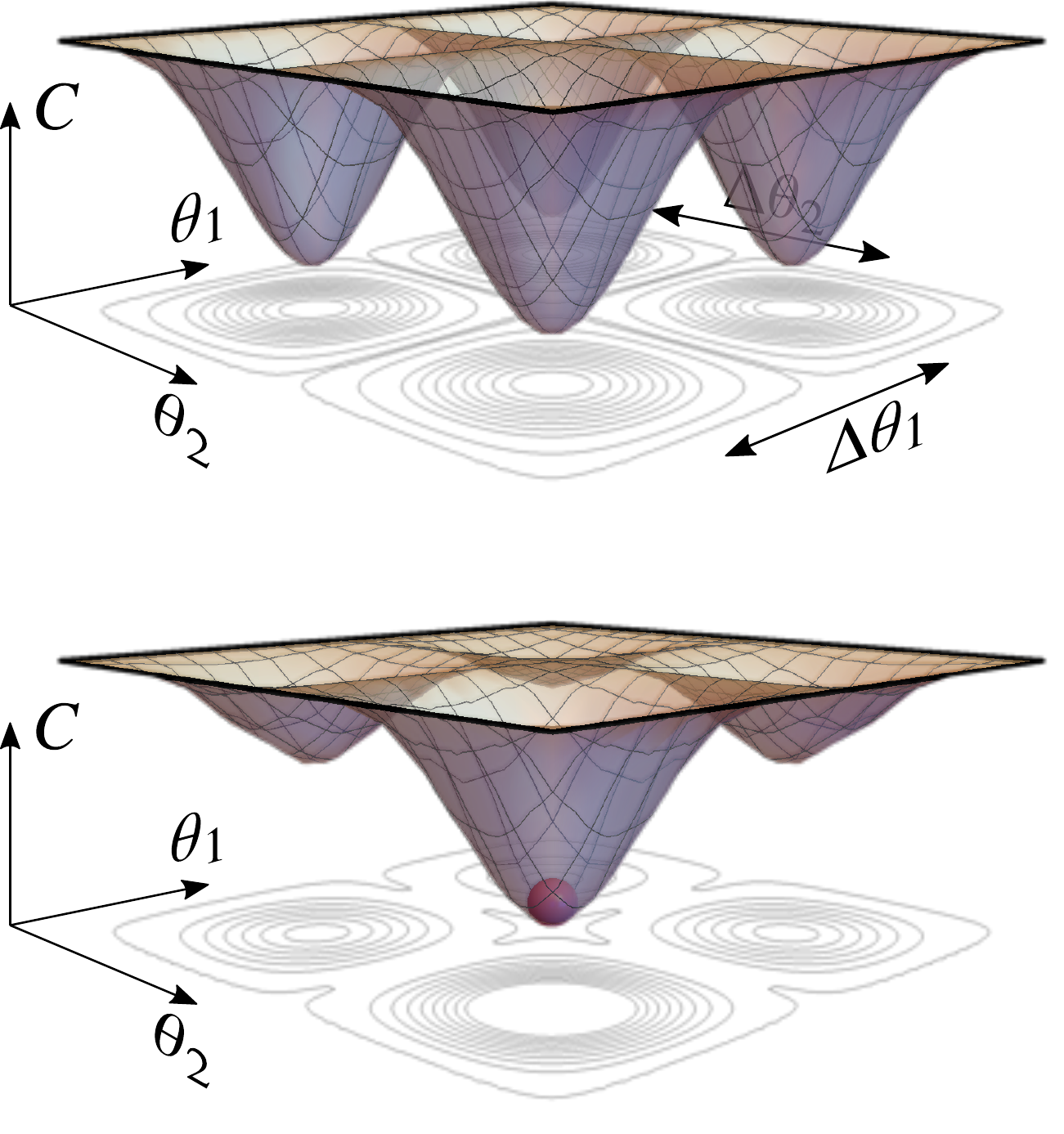}
    \caption{(a) Schematic diagram of a noise-free cost landscape. Here the cost function is periodic under some translation in the parameters of the ansatz $V(\thv)$, which leads to the global minima being degenerate. By randomly initializing and optimizing $\thv$ one can converge to any of the degenerate minima.  (b)~Schematic diagram of the noisy cost landscape. The presence of quantum noise breaks the periodicity in the landscape, leading to some of the global minima that were degenerate in the noiseless scenario to become local minima. Now, when optimizing $\thv$ one can get trapped in a local minimum.  }
    \label{fig:Landscapes}
\end{figure}

At the heart of our work is a novel understanding of symmetries and symmetry breaking in PQCs. In particular, we analytically prove that given some non-restrictive conditions on the ansatz, the cost landscape must have a certain periodicity, which gives rise to a large degeneracy of minima in the absence of noise. This is schematically shown in Fig.~\ref{fig:Landscapes}(a). In the noise-free setting, one can move between these degenerate minima with pulses that rotate the parameters by specific angles. 

However, our second analytical result is that the symmetry can be broken by certain types of noise, specifically, non-unital noise (such as amplitude damping) and coherent noise. As a consequence, the degeneracy of various minima is lifted by noise, leading to landscapes like in Fig.~\ref{fig:Landscapes}(b). We denote this phenomenon noise-induced breaking of symmetries (NIBS). 

The NIBS phenomenon, in turn, allows us to construct an optimizer where one exploits circuit symmetries to hop between local minima valleys. Because these valleys are no longer degenerate, such hopping potentially leads to lower cost values and mitigate the effect of noise by leading to better solutions.  This is the idea behind our Symmetry-based Minima Hopping technique. 

In what follows, we first lay out our general framework in the next section. We then discuss symmetries in PQCs in Section~\ref{sec:symmetries}. In Section~\ref{sec:Pulse} we present a method to move between these symmetries that we call the $\sigma$-Pulse method. Section~\ref{sec:NIBS} considers the impact of noise and presents our results on the NIBS phenomenon. Section~\ref{sec:SYMH} presents our SYMH optimizer. Finally, Section~\ref{sec:numerics} shows our numerical implementations of SYMH, and demonstrates that SYMH leads to a significant improvement in the optimizer performance for various problems under realistic noise models.

\section{General Framework}

In this section we introduce the general framework for our work. Specifically, we consider here a generic quantum machine learning task where the goal is to minimize a parametrized cost function of the form
\begin{equation}~\label{eq:cost}
    C(\thv)=\sum_{x=1}^{S} f_x(\rho_x,V(\thv))\,.
\end{equation}
Here, $\SC=\{\rho_x\}_{x=1}^S$ is a training set of input states,  and  $f_x$ is a function that determines the task at hand and which can  be different for each input state. Moreover,  $V(\thv)$ is a PQC, and $\thv$ are trainable parameters. 

The cost function in~\eqref{eq:cost} is in fact quite general and includes as special cases the cost functions for many important VQAs and QNNs. For example, in the Variational Quantum Eigensolver~\cite{VQE} we have $S=1$ and the cost $C=\Tr[HV(\thv)\rho V\ad(\thv)]$ is  the expectation value of a given Hamiltonian $H$. Alternatively, in a binary classification problem, the cost can be expressed as  the mean-squared error $C=\frac{1}{2K}\sum_x[y_x-\widetilde{y}(\rho_x,V(\thv))]^2$, with $y_x$ the true label, and $\widetilde{y}_x(\rho_x,V(\thv))$ the predicted label for each state in the training set~\cite{cong2019quantum}.

We here consider PQCs $V(\thv)$  that can be expressed as the product of $L$ unitaries as
\begin{equation}\label{eq:V}
    V(\thv)=V_L(\thv_L)\cdots V_1(\thv_1)\,,
\end{equation}
where $\thv=\{\thv_l\}_{l=1}^L$ is a set of continuous parameters. Each unitary $V_l(\thv_l)$ can in turn be expanded as
\begin{equation}\label{eq:Vl} 
     V_l(\thv_l)=\prod_{m_l=1}^{\eta_l}R_\mu(\theta_{m,l})W_{m_l}\,,
\end{equation}
where  $R_\mu(\theta)=e^{-i\sigma_{\mu}\theta_{m_l}}$ is a single qubit rotation, and where $\sigma_{\mu}\in\{X_j,Y_j,Z_j\}_{j=1}^{n}$ is a Pauli operator on qubit~$j$. Moreover, $W_{m_l}$ denote  unparametrized gates. For simplicity, we consider here that $W_{m_l}$  are CNOTs or identities. That is, $W_{m_l}\in\{\id_i,C_X^{ij}\}_{i,j\in\CC}$, where $C_X^{ij}$ is a CNOT with control qubit $i$ and target qubit $j$, and where $\CC$ is a graph of the qubit connectivity. Finally, take $M$ to be the total number of controllable parameters in the PQC.

The PQC in~\eqref{eq:V}--\eqref{eq:Vl} includes as special cases many ansatzes widely used in the literature. For instance, if $V(\thv)$ is a  hardware-efficient ansatz~\cite{kandala2017hardware}, then the set of available CNOTs is determined by the device connectivity. In addition, $V(\thv)$ also includes as special cases the Quantum Alternating Operator Ansatz (QAOA)~\cite{qaoa2014,nasaQAOA2019}, and the Unitary Coupled Cluster (UCC)~\cite{cao2019quantum,bartlett2007coupled,lee2018generalized} ansatz.  Specifically, when implementing the QAOA or a UCC ansatz one usually performs first order Trotter approximations of unitaries of the form $e^{-i\theta H}$, where $H$ is a Hermitian operator with an efficient decomposition in the Pauli basis.  The latter then leads to a PQC that fits into the framework presently considered~\cite{wang2020noise}. 

\begin{figure}[t]
    \centering
    \includegraphics[width=.95\columnwidth]{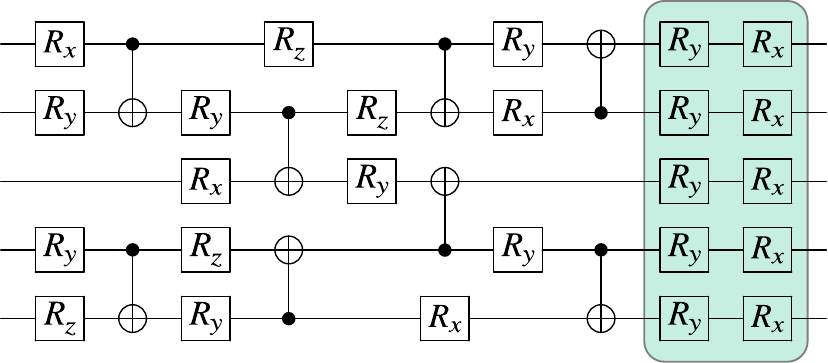}
    \caption{Schematic representation of a buffered PQC. The buffered version of any parametrized quantum circuit $V(\thv)$ can be obtained by appending to the circuit a buffer unitary as described in~\eqref{eq:buffer}. Here we have highlighted this buffer unitary.   }
    \label{fig:Ansatz}
\end{figure}

Given a PQC $V(\thv)$, we define its {\it buffered} version as follows.
\begin{definition}[Buffered PQC]\label{def:1}
Let $V(\thv)$ be a PQC as in~\eqref{eq:V} and~\eqref{eq:Vl}. We define the buffered version of this PQC as a gate sequence 
\begin{equation}
    V_B(\thv,\gamv)=U_B(\gamv) V(\thv)\,,
\end{equation}
where $U_B(\gamma)$ is the so-called {\it buffer} unitary given by
\begin{equation}\label{eq:buffer}
    U_B(\gamv)=\bigotimes_{j=1}^n e^{-i X_j \gamma_{2j}}e^{-i Y_j \gamma_{2j-1}}\,.
\end{equation}
\end{definition}
As shown in Fig.~\ref{fig:Ansatz}, the buffer unitary is simply given by a tensor product of single qubit rotations around the $x$ and $y$ axes which are parametrized by the vector $\gamv$ of length $2n$. 

We note that our main results, stated below, are valid for buffered PQCs. However, since $V(\thv)=V_B(\thv,\vec{0})$ one can always trivially extend any PQC to its buffered version. That is, any PQC of the form~\eqref{eq:V}--\eqref{eq:Vl} can be considered as a buffered PQC with trivial rotation angles in the buffer unitary. In addition, we also remark that our results will also hold for any PQC where $V(\thv)$ contains (at least) two single-qubit rotations about different axes in every qubit (not necessarily sequentially or in parallel). However, for the sake of simplicity in introducing the method, we consider the case where one appends a buffer unitary to the PQC.

Finally, we remark that in some practical settings the buffer unitary can be implemented virtually without any additional computational overhead. Whenever $V(\thv)$ acts before measurement, the single-qubit rotations in the buffer layer can be absorbed into the measurement operator and executed classically by post-processing the measurement statistics. However, when the buffered unitary does not act prior to the measurements, such as when only a portion of the circuit is buffered, then $U_B(\gamv)$ must be included.

\section{$\sigma$-Pulse method for finding parameter symmetries in PQCs}\label{sec:Pulse2}

In this section, we first discuss symmetries in PQCs, which lead to degenerancies in the cost function landscape. We then present a method to move between different sets of parameters that are symmetric by propagating Pauli gates throughout the circuit, which we therefore call the $\sigma$-Pulse method.

\subsection{Symmetries in PQCs}\label{sec:symmetries}

As it was recently pointed out in~\cite{fontana2020evaluating}, different sets of parameters $\thv$ in a PQC can leads to the same unitary being produced. In order to further analyze this phenomenon, we first introduce the following definition:
\begin{definition}[Parameter symmetries]\label{def:param_degs}
Let $V(\thv)$ be a PQC. We say that two distinct sets of parameters $\thv$ and $\widetilde{\thv}$ are symmetric if $V(\thv)$ is equal to $V(\widetilde{\thv})$ (up to a global phase).
\end{definition}
Let us here make two important remarks.
First, note that Definition~\eqref{def:param_degs} implies that the structure of the circuit remains unchanged between $V(\thv)$ and $V(\widetilde{\thv})$, as no gates in the circuit are being added or replaced; only their parameter values differ. 
Second, we remark that these parameter symmetries naturally translate into cost function landscape degenerancies. That is, given two symmetric sets of parameters $\thv$ and $\widetilde{\thv}$ we have  $C(\thv)=C(\widetilde{\thv})$.

Note that there are many mechanics which can lead to symmetries in $\thv$. For instance, they can arise from the wrapping symmetry in a rotation, i.e., from the fact that for any single qubit rotation we have   $R_\mu(\theta_\mu)=R_\mu(\theta_\mu+2\pi)$. Similarly,  parameter symmetries can also be obtained from other types of mechanisms, such as commutation symmetries. Consider for example a two-qubit PQC composed of a CNOT preceded and followed by single qubit rotations about the $z$ axis on the first qubit. That is, $V(\theta_1,\theta_2)=\left(R_z(\theta_1)\otimes\id_2\right).C_X^{12}.\left(R_z(\theta_2)\otimes\id_2\right)$. Noting that  $[\left(R_z(\theta_1)\otimes\id_2\right),C_X^{12}]=0$, then it follows that $V(\theta_1,\theta_2)=V(p\theta_1+(1-q)\theta_2,(1-p)\theta_1+q\theta_2)$ for any $\theta_1$ and $\theta_2$, and for any $p,q\in(0,1)$. 

In the work we provide a general theory to analyze a class of non-trivial discrete parameter symmetries which were first reported in~\cite{fontana2020evaluating} and which we generalize here. Specifically,  we introduce a method for finding and characterizing the following symmetries:
\begin{definition}[$\sigma$-Pulse symmetries]\label{def:sym}
Let $V_B(\thv,\gamv)$ be a buffered PQC as in Definition~\ref{def:1}. $V_B(\thv,\gamv)$ possesses a $\sigma$-Pulse symmetry whenever, for any $\{\thv,\gamv\}$, there exists a set of $\{\widetilde{\thv},\widetilde{\gamv}\}$ related to $\{\thv,\gamv\}$ via
\begin{equation}\label{eq:restriction}
    \widetilde{\theta}_j=(-1)^{p_j}\theta_j+{q_j}\pi\,,\quad  \widetilde{\gamma}_j=(-1)^{p_j'}\gamma_j+{q_j'}\pi\,,
\end{equation}  
for some $p_j,q_j,p_j',q_j'\in\{0,1\}$, such that $\{\thv,\gamv\}$ and $\{\widetilde{\thv},\widetilde{\gamv}\}$  are symmetric. 
\end{definition}

In the next section we present the so-called $\sigma$-Pulse method for determining $\sigma$-Pulse symmetries.

\subsection{$\sigma$-Pulse method}\label{sec:Pulse}

Here we introduce the $\sigma$-Pulse method which allow us to start from a set of parameters $\{\thv,\gamv\}$ of a buffered PQC and obtain a second set  $\{\widetilde{\thv},\widetilde{\gamv}\}$ which are $\sigma$-Pulse symmetric to $\{\thv,\gamv\}$ according to Definition~\ref{def:sym}. This method is based on three basic steps: (1) The creation of the so-called $\sigma$-Pulses, (2) The propagation of said pulses through the circuit, and (3) The absorption of the $\sigma$-Pulses in the buffer unitary. The mathematical rules for these steps (described below in detail) are schematically shown in the $ZX$-calculus~\cite{coecke2011interacting} notation of Fig.~\ref{fig:sigmapulse}(a).

\subsubsection{Creation of $\sigma$-Pulses}

The first step of the $\sigma$-Pulse method is based on the fact that any single qubit rotation around a principal axis satisfies the following identity
\begin{align}\label{eq:pipulse}
    R_\mu(\theta)&=e^{-i\theta \sigma_\mu/2}= e^{-i(\theta+\pi) \sigma_\mu/2} e^{i\pi \sigma_\mu/2} \nonumber\\
    &= R_\mu(\theta+\pi) \left( i\sigma_\mu\right)=  R_\mu^{01}\left( i\sigma_\mu\right)\,,
\end{align}
where we defined the shifted rotations
\begin{equation}\label{eq:shifted}
    R_\mu^{pq}=R_\mu\left((-1)^p\theta+q\pi\right)\,,
\end{equation} 
for $p,q\in\{0,1\}$ and $\mu \in \{x, y, z\}$. As shown in Fig.~\ref{fig:sigmapulse}(b),  Eq.~\eqref{eq:pipulse} implies that the angle of a single-qubit rotation in $V(\thv)$ can be shifted by $\pi$ at the expense of adding to the circuit a $i \sigma_\mu$ gate, i.e., at the expense of creating a  $\sigma$-Pulse. When Eq.~\eqref{eq:pipulse} is employed to generate a pulse, we say that the gate is a {\it generator} of a {\it primary} pulse. Note that simply employing~\eqref{eq:pipulse} changes the structure of the ansatz as we have added new gates to the circuit. For the structure of $V_B(\thv,\gamv)$ to be preserved, the $\sigma$-Pulses need to be propagated through the circuit towards $U_B(\gamv)$ where they be absorbed.

\subsubsection{Propagation of $\sigma$-Pulses}

\begin{figure}[t]
    \centering
    \includegraphics[width=.95\columnwidth]{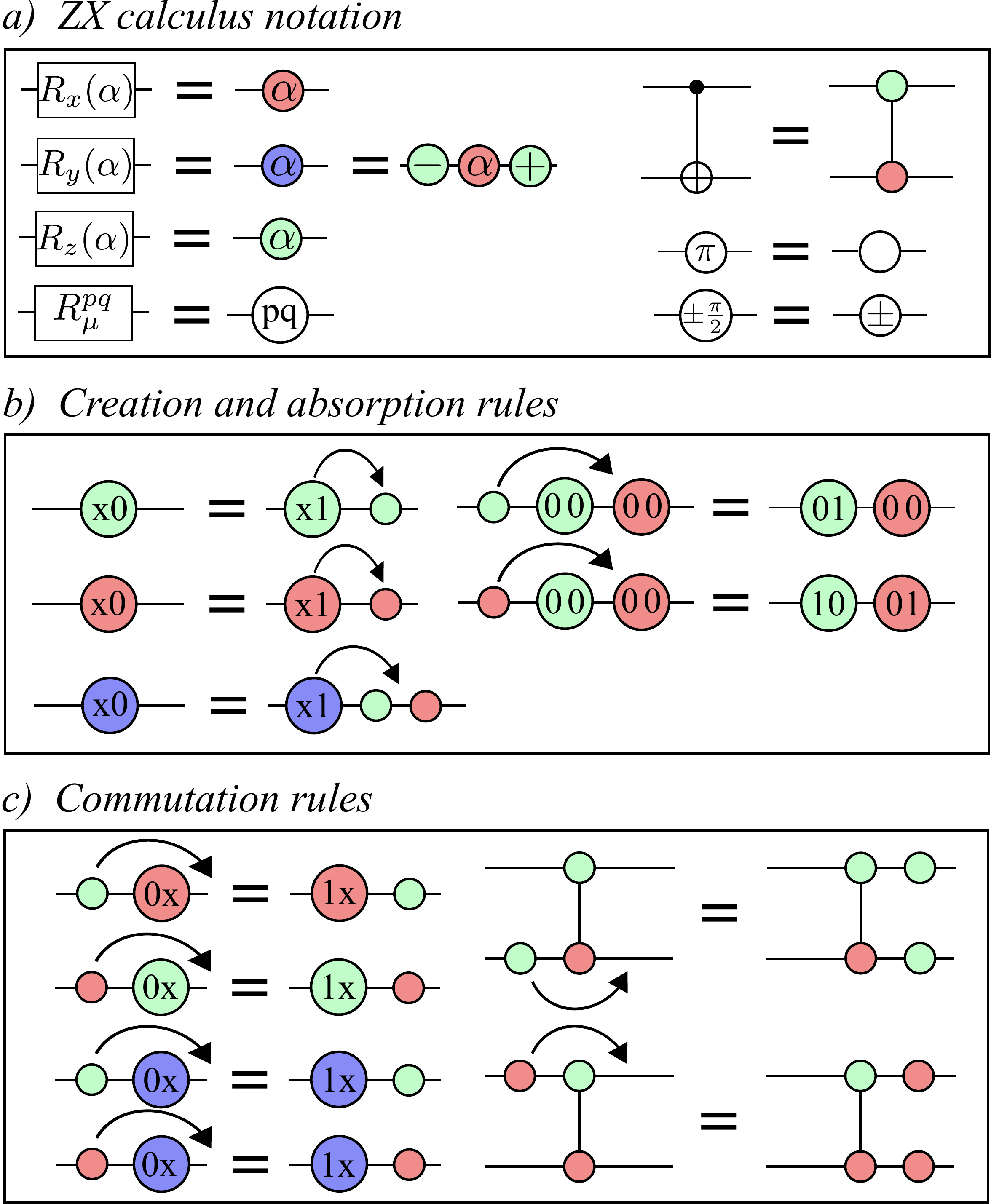}
    \caption{$\sigma$-Pulse method rules in the notation of ZX-calculus. (a) Notation for  rotations around the principal axes, shifted rotations (of Eq.~\eqref{eq:shifted}), and for a CNOT gate. We additionally introduce notation for the $\sigma$-Pulses (rotation of angle $\pi$), and for rotations of angles $\pm \frac{\pi}{2}$. (b) Rules for creating and absorbing $\sigma$-pulses.  (c) Non-trivial commutation rules for $\sigma$-Pulses. We remark that in panels (b) and (c), the rules fora $(i\sigma_y)$ pulse can be derived from those of $(i\sigma_x)$ and $(i\sigma_z)$.  In addition, in those panels an equal sign indicates that the unitaries are equal up to a global phase.  }
    \label{fig:sigmapulse}
\end{figure}

Propagating the $\sigma$-Pulses through the circuit implies knowing how  $i\sigma_\mu$ commutes with all other gates in the ansatz.  The commutation of a $\sigma_\mu$-Pulse through a single qubit rotation $R_\nu(\theta)$ is given by
\begin{equation}\label{eq:pipulserot}
         R_\nu (\theta) \left(i\sigma_\mu\right) = i\sigma_\mu R_\nu\left((-1)^{\delta_{\mu\nu}}\theta\right)= i\sigma_\mu R_\nu^{\delta_{\mu\nu}0}\,.
\end{equation}
Equation~\eqref{eq:pipulserot} shows that if $\mu\neq \nu$, the commutation of a pulse with a rotation can lead to said rotation picking up a minus sign. Moreover, the following identities provide the commutation rules between a pulse and a CNOT:
\begin{equation}\label{eq:pipulsecx}
\begin{aligned}
      C_X   (i\sigma_\mu\otimes \id) &=  -i (i\sigma_\mu\otimes iX)  C_X,\,\,\,\text{for }\sigma_\mu=Y,X\\
      C_X   (\id\otimes i\sigma_\mu) &= -i (iZ\otimes i\sigma_\mu)  C_X,\,\,\,\,\text{for }\sigma_\mu=Y,Z\\ 
      [C_X,   (iZ\otimes \id)] &=[C_X,   (\id\otimes iX)] =0\,,
\end{aligned}
\end{equation}
These commutation rules,  which we illustrate in Fig.~\ref{fig:sigmapulse}(c), show that commuting a  $\sigma$-Pulse on the control (target) qubit through a CNOT can lead to the creation of a {\it secondary} pulse on the target (control) qubit, plus a global unobservable phase. For the gate structure of $V_B(\thv,\gamv)$ to remain unchanged, the secondary pulses also need to be propagated towards the measurement, which in turn means that they can create additional secondary pulses.

\subsubsection{Absorption of $\sigma$-Pulses}

Once all  the primary and secondary pulses have been propagated to the buffer unitary, they can be absorbed by shifting the rotation in $U_B(\gamv)$ angles as 
\begin{align}
R_x(\gamma') R_y(\gamma)(i X )&= R_x(\gamma'-\pi) R_y(-\gamma)=-R_x^{01}R_y^{10},\nonumber\\
R_x(\gamma') R_y(\gamma)(i Y) &= R_x(\gamma') R_y(\gamma-\pi)=-R_x^{00}R_y^{01},\label{eq:buffer2}\\
R_x(\gamma') R_y(\gamma) (iZ) &= R_x(\gamma'-\pi) R_y(\pi-\gamma)=-R_x^{01}R_y^{11}\nonumber,
\end{align}
where we use the definition of the shifted rotations $R_\mu^{pq}$ of~\eqref{eq:shifted}. Here we remark that the minus signs on the right-hand side of~\eqref{eq:buffer2} simply correspond to unobservable global phases. These absorption rules are shown in Fig.~\ref{fig:sigmapulse}(b).

\subsubsection{Parameter symmetries}

\begin{figure}[t]
    \centering
    \includegraphics[width=.85\columnwidth]{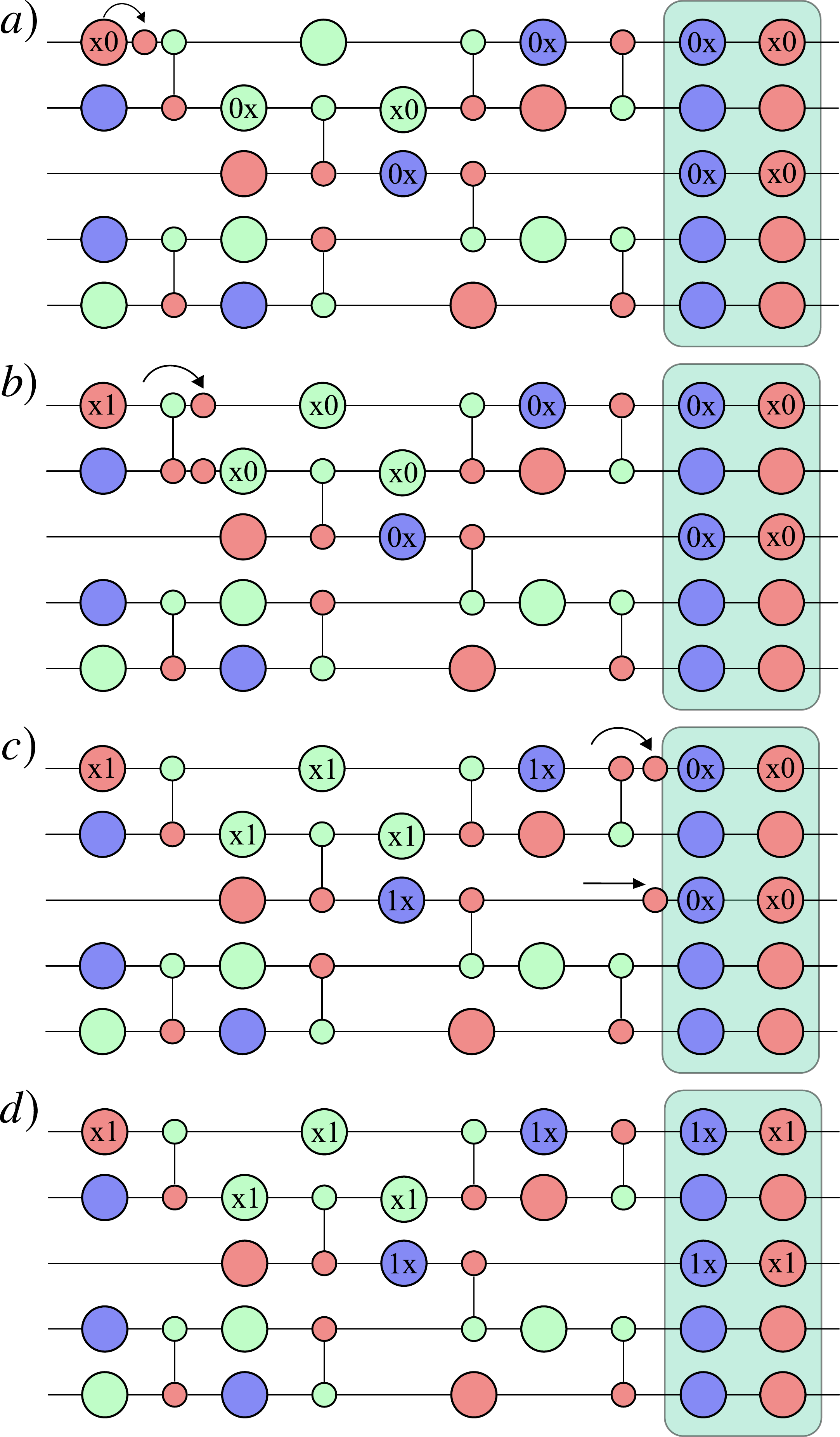}
    \caption{Schematic of the $\sigma$-Pulse method for finding parameter symmetries. We start with the buffered PQC of Fig.~\ref{fig:Ansatz}(a) with parameters $\{\thv,\gamv\}$. (a)   A primary $\sigma$-Pulse is generated in the first $R_x$ rotation acting on the first qubit. (b) While propagating the pulse through the CNOT gate, secondary pulses are created. (c) All the primary and secondary pulses are propagated to the buffer layer, where they can be absorbed. (d) Once the pulses are absorbed, we have a new set of parameters $\{\widetilde{\thv},\widetilde{\gamv}\}$ which are symmetric to $\{\thv,\gamv\}$ according the Definition~\ref{def:sym}. }
    \label{fig:degenerancies}
\end{figure}

Equations~\eqref{eq:pipulserot}--\eqref{eq:buffer2} provide the framework for determining symmetries in $V_B(\thv,\gamv)$ with the $\sigma$-Pulse method. Given a set of angles $\{\thv,\gamv\}$, one can select any number of rotations to generate primary pulses. Once the primary and secondary pulses are propagated and absorbed in the buffer unitary, we define $\{\widetilde{\thv},\widetilde{\gamv}\}$ as the ensuing new set of angles. From Eqs.~\eqref{eq:pipulse} and~\eqref{eq:buffer2} it is straightforward to see that $\{\widetilde{\thv},\widetilde{\gamv}\}$ and $\{\thv,\gamv\}$ are symmetric according to Eq~\eqref{eq:restriction} in  Definition~\ref{def:sym}. In Fig.~\ref{fig:degenerancies} we explicitly show this procedure.

Definition~\eqref{def:sym}, and more specifically, Eq.~\eqref{eq:restriction}, allows us to derive the following proposition, which is  proved in the Appendix. 
\begin{proposition}[Exponential Symmetry]
\label{prop1}
Let $V_B(\thv,\gamv)$ be a buffered PQC. Then, for any $\{\thv,\gamv\}$ there exists $2^{M}$ sets of $\sigma$-Pulse symmetric parameters $\{\widetilde{\thv},\widetilde{\gamv}\}$ according to Definition~\eqref{def:sym}. Each symmetric set can be characterized by a bitstring $\vec{\beta}$ of length $M$ such that  $\beta_j=0$ if  $\widetilde{\theta}_j\in [0,\pi)$ and $\beta_j=1$ if  $\widetilde{\theta_j}\in [\pi,2\pi)$. We denote as $\BC$ the set of such bitstrings. 
\end{proposition}

Proposition~\ref{prop1} has several important implications. First, it shows that the each point in the cost function landscape is exponentially periodic, as the cost is symmetric over the parameter translation $\theta_j\rightarrow (\theta_j+\pi)$ for every $j$. This holds up to a sign change $\theta_k \rightarrow -\theta_k$ in some other parameters with $k\neq j$ and a correction rotation in the buffer layer parameters $\gamv$, due to the necessity to propagate and absorb the extra $\sigma$-Pulses. In particular, denoting as $\{\thv_\opt,\gamv_\opt\}$ a set of parameters that minimize  the cost $C(\thv,\gamv)$ we have that the global minimum is $2^M$-fold degenerate. 

In addition, Proposition~\ref{prop1} implies  the next Corollary. 
\begin{corollary}[Effective Parameter Space Reduction]
\label{cor:reduction}
Let $V_B(\thv,\gamv)$ be a buffered PQC . Then, for every $\{\thv,\gamv\}$  there always exists a set of parameters $\{\widetilde{\thv},\widetilde{\gamv}\}$ which are $\sigma$-Pulse symmetric to $\{\thv,\gamv\}$, and which are such that $\widetilde{\theta_j}\in[0,\pi)$ for all $j$.
\end{corollary}
The proof of Corollary~\eqref{cor:reduction} follows from Proposition~\ref{prop1} by taking $\vec{\beta}=\vec{0}$. This Corollary implies an exponential reduction of the effective hyperarameter space by restricting the domain of all angles in  $V(\thv)$ from  $[0,2\pi)^{M}$ to $[0,\pi)^{M}$. Hence, all relevant features  of the cost function landscape (including the global minima) can be found in $[0,\pi)^{M}$. We finally remark that the domain restriction in Corollary~\eqref{cor:reduction} is non-trivial as it does not arise from a wrapping symmetry in the rotation parameters (i.e., it does not arise from the fact that $R_{\mu}(\theta)=R_{\mu}(\theta+2\pi)$). Instead this domain reduction arises from the $\sigma$-Pulse symmetries.

\section{Noise-induced lifting of the symmetries}\label{sec:NIBS}

In this section we analyze how noise affects the symmetries in $V_B(\thv,\gamv)$ and hence the degenerancies in the  cost landscape.  Our main results are presented in the form of two theorems, with Theorem~\ref{theo:1} analyzing the effect of unital Pauli noise, and Theorem~\ref{theo:2} the effect of non-unital Pauli noise. We recall that unital Pauli noise channels include $T2$ processes (i.e. dephasing channel), and depolarizing  as special cases. On the other hand, non-unital Pauli noise channels include $T1$ processes as a special case, i.e. the amplitude damping channel is a non-unital Pauli channel.

\subsection{Unital Pauli noise}

Consider the following definition: 
\begin{definition}[Unital Pauli noise model]\label{def:Unital}
We define the unital Pauli noise model as a process in which a unital Pauli channel acts after every layer of gates acting in parallel in $V_B(\thv,\gamv)$.
\end{definition}
Here we recall that unital Pauli noise channels are completely positive trace-preserving maps $\mathcal{P}_{U}$ whose superoperator is diagonal in the Pauli basis. The action of $\mathcal{P}_{U}$ on an $n$-qubit Pauli operator $X^{\vec{a}}Z^{\vec{b}}$ is given by 
\begin{align}
\mathcal{P}_{U}(X^{\vec{a}}Z^{\vec{b}})&= p_{\vec{a}\vec{b}}X^{\vec{a}}Z^{\vec{b}} \,,\label{eq:unitalPauli}
\end{align}
 where $p_{\vec{0}\vec{0}}= 1$, and where we assume that $-1\leq p_{\vec{a}\vec{b}}\leq 1$ $\forall \vec{a},\vec{b}$.
Here $\vec{a},\vec{b}\in\{0,1\}^{\otimes n}$ are bitstrings of length $n$, and where  we employ the notation
\begin{equation}
X^{\vec{a}}= X^{a_1}_1\otimes\dotsb\otimes X^{a_n}_n\,,\qquad Z^{\vec{b}}= Z^{b_1}_1\otimes\dotsb\otimes Z^{b_n}_n\,.
\end{equation}

As explicitly shown in the Appendix, the following theorem holds.
\begin{theorem}[Symmetry preservation]
\label{theo:1}
Let $V_B(\thv,\gamv)$ be a buffered PQC  as in Definition~\ref{def:1}. Then, the $\sigma$-Pulse parameter symmetries in $V_B(\thv,\gamv)$ are preserved under the action of the unital Pauli noise model in Definition~\ref{def:Unital}. 
\end{theorem}
Theorem~\ref{theo:1} shows that the parameter symmetries in  $V_B(\thv,\gamv)$ arising from $\sigma$-Pulse symmetries are completely preserved by the action of unital Pauli noise channels which includes as special cases the action of local (or global) depolarizing channels, as well as dephasing channels. 

In addition, Theorem~\ref{theo:1} implies that the degeneracy in the cost function landscape also remains unchanged. Particularly, we then know that the optimal parameters $\{\widehat{\theta}_\opt,\widehat{\gamv}_\opt\}$ leading to the global minimum of the noisy cost function will still be $2^M$-fold degenerate in $[0,2\pi)$. That is, starting from $\{\widehat{\theta}_\opt,\widehat{\gamv}_\opt\}$, all the symmetric parameters obtained from any bitstring in $\BC$ will have the same energy. In a practical scenario Theorem~\ref{theo:1}  implies that the minimum cost achievable from randomly initializing the parameters $\{\thv,\gamv\}$  will be independent of the bitstring $\vec{\beta}$ characterizing the initial point. 

Note however that for general cost functions the presence of quantum noise can change the cost landscape such that the optimal parameters  of the noisy cost function are different than the ones for the noiseless case~\cite{mcclean2016theory}. That is, one generally has  $\{\widehat{\theta}_\opt,\widehat{\gamv}_\opt\}\neq \{\theta_\opt,\gamv_\opt\}$. For the special cases when $\{\widehat{\theta}_\opt,\widehat{\gamv}_\opt\}= \{\theta_\opt,\gamv_\opt\}$ we say that the cost has {\it optimal parameter resilience}. This phenomenon has been analyzed in~\cite{sharma2019noise} for the problem of variational quantum compiling. 

In addition, we also know that the value of the noisy cost function evaluated at the optimal parameters $\widehat{C}(\widehat{\theta}_\opt,\widehat{\gamv}_\opt)$ can also change due to the presence of noise. Here $\widehat{C}$ denotes the noisy cost function. In fact, let us consider a cost function of the form $C(\thv,\gamv)=\Tr[OV(\thv)\rho V\ad(\thv)]$, where $O$ is a Hermitian operator, and assume that the parameters $\{\theta_\opt,\gamv_\opt\}$ yield the operator's ground state. Then we have trivially that
\begin{equation}
    C(\thv_\opt,\gamv_\opt)\leq \widehat{C}(\widehat{\thv}_\opt,\widehat{\gamv}_\opt)\,.
\end{equation}
In general, the output state of the PQC converges to the fixed point of the noise model~\cite{franca2020limitations}, and so the cost function will be increasingly different as noise increases.

\subsection{Non-unital Pauli noise}

Let us here analyze the effect of non-unital Pauli noise on the $\sigma$-Pulse symmetries. Hence, consider the following definition 
\begin{definition}[Non-unital Pauli noise model]
We define the non-unital Pauli noise model as a process in which a non-unital Pauli channel acts after every layer of gates acting in parallel in $V_B(\thv,\gamv)$.  
\end{definition}
Here we recall that non-unital Pauli noise channels are completely positive trace-preserving maps $\mathcal{P}_{NU}$ whose action on the identity operator is 
\begin{equation}
    \mathcal{P}_{\text{NU}}(\id)=\id+ \sum_{(\vec{a},\vec{b})\neq (\vec{0}, \vec{0})}d_{\vec{a}\vec{b}}X^{\vec{a}}Z^{\vec{b}}\,.\label{def:non-un-Puali2}
\end{equation}
On the other hand, its action on all other Pauli operators is given by
\begin{align}
\mathcal{P}_{\text{NU}}(X^{\vec{a}}Z^{\vec{b}})= q_{\vec{a}\vec{b}}X^{\vec{a}}Z^{\vec{b}} \,.\label{def:non-un-Puali1}
\end{align}

As explicitly shown in the Appendix, the following theorem holds.
\begin{theorem}[Symmetry breaking]
\label{theo:2}
Let $V_B(\thv,\gamv)$ be a buffered PQC  as in Definition~\ref{def:1}. Then, the $\sigma$-Pulse parameter symmetries in $V_B(\thv,\gamv)$ can be broken under the action of a non-unital Pauli noise model.
\end{theorem}

From Theorem~\ref{theo:2} we have that given two sets of symmetric parameters $\{\thv,\gamv\}$ and $\{\widetilde{\thv},\widetilde{\gamv}\}$, then there exists some non-unital noise such that we have $V(\thv,\gamv)\neq V(\widetilde{\thv},\widetilde{\gamv})$. This implies that the $2^M$-fold symmetry of the optimal noisy parameters, and hence the degeneracy in the cost landscape, is broken. 

In particular we now have that some of the (previously) global minima are transformed into local minima. In practical terms, due to this degeneracy breaking in the cost landscape not all randomly initialized $\{\thv,\gamv\}$ will converge to the global minima. In the next section we show how this effect can be mitigated by exploiting the knowledge of parameter symmetries to construct am optimizer.

We remark that the proof for Theorem~\ref{theo:2} in the Appendix is valid for more general noisy channels that include as special case non-unital Pauli nose channels. For instance, we show that coherent error, such as qubit drift, can also break the $\sigma$-Pulse symmetries. 

\section{Symmetry-based Minima Hopping (SYMH) optimizer}\label{sec:SYMH}

Here we present the Symmetry-based Minima Hopping (SYMH) optimizer, which is meant to be employed in the presence of non-unital quantum noise. As its name indicates SYMH employs the $\sigma$-Pulse symmetries to hop around the degeneracy-broken landscape and attempt to find the minima that are less sensitive to noise. As further explained below, the strength of SYMH is that it should be considered a general tool which can be implemented along with other optimization and error mitigation techniques.

\begin{figure}
    \centering
    \includegraphics[width=0.4\textwidth]{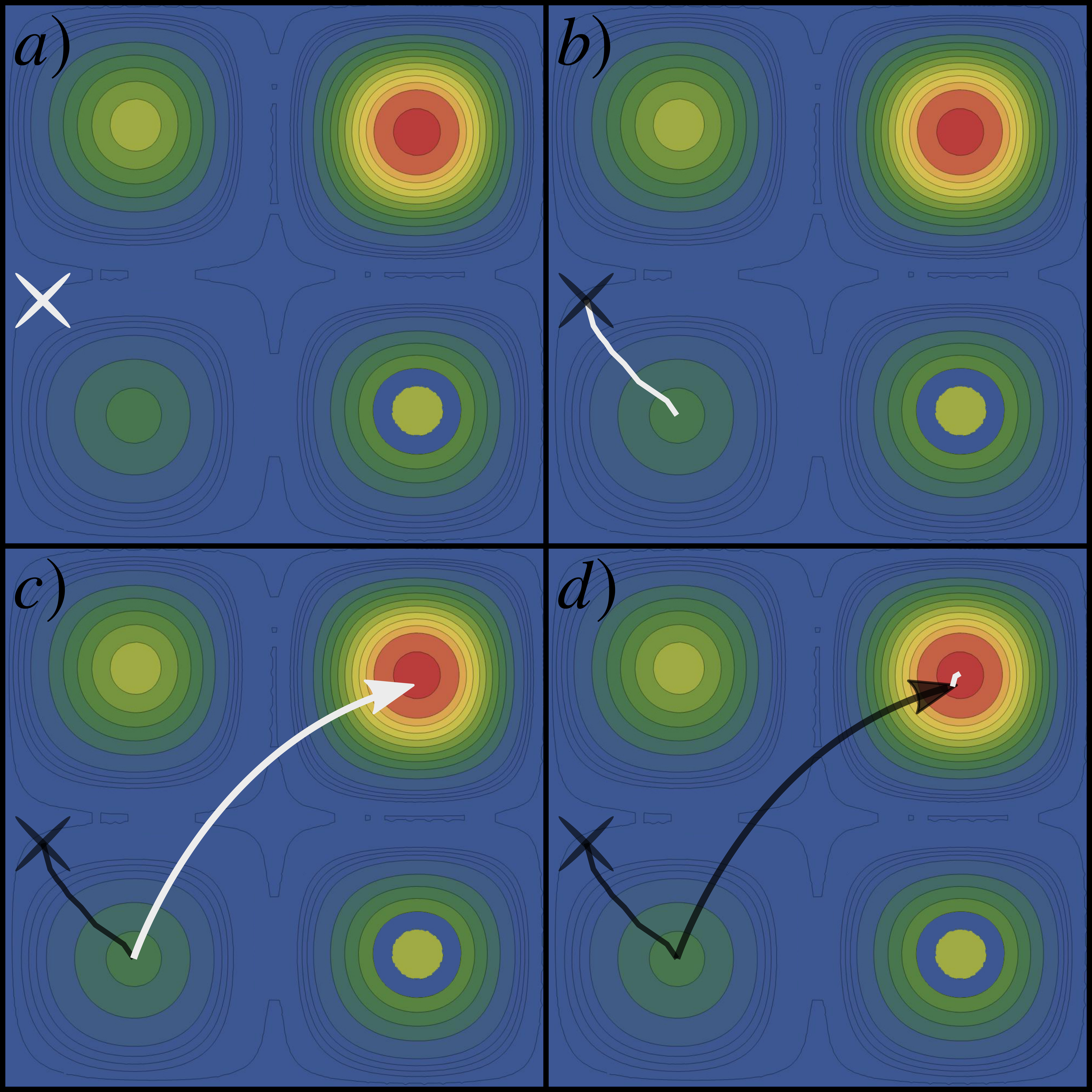}
    \caption{Main idea behind the SYMH optimizer. The contour plots correspond to the cost landscape of Fig.~\ref{fig:Landscapes}(b), where the presence of non-unital quantum noise has broken the landscape degeneracy. a) The parameters are random initialized. b) By employing a classical optimizer we can determine a cost minimizing direction. In this case, the minima to which the optimizer converges is a local minima which used to correspond to a global minima in the noiseless case. c) Using the $\sigma$-Pulse method we can hop in the landscape and land in the vicinity of another minima. d) By performing an optional second optimization we can find the global minima of the problem.}
    \label{fig:SYMH}
\end{figure}

Consider the problem of minimizing the cost $C(\thv,\gamv)$ function in Eq.~\eqref{eq:cost}. As shown in Fig.~\ref{fig:SYMH}(a), a common strategy is to randomly initialize the parameters $\{\thv,\gamv\}$ and employ a classical optimizer which takes as input the value of the cost (or its gradients) to determine a cost minimizing direction. Usually one optimizes until some stopping criteria has been met, at which point one hopes that the minima reached corresponds to a global optima. We call the parameters obtained at the end of the optimization as $\{\thv_f,\gamv_f\}$, where the $f$ stands for final. 

However, as shown in Fig.~\ref{fig:SYMH}(b),  randomly initializing the parameters can lead to the optimizer getting trapped in local minimum which used to correspond to global minimum in the noiseless scenario. In this case, one can attempt to find the global minimum by taking final parameters $\{\thv_f,\gamv_f\}$ and employing the $\sigma$-Pulse method (i.e. using Eq.~\eqref{eq:restriction}) to obtain a new set of parameters   $\{\widetilde{\thv}_f,\widetilde{\gamv}_f\}$. As depicted  in Fig.~\ref{fig:SYMH}(c) this will effectively lead to a hop in the cost-landscape whose ending point can be in the well of another minima. 

Since the cost landscape degeneracy is broken, then $\{\widetilde{\thv}_f,\widetilde{\gamv}_f\}$ might not correspond to a critical point: one therefore has to re-optimize. This is schematically shown in Fig.~\ref{fig:SYMH}(d), where the hop leads to the vicinity of the minimum and additional optional optimization could be needed. Note that in general this optional optimization will only require a small number of circuit evaluations, and hence will not add significant overhead to the optimization. 

Here we remark that there are several possibilities for where the SYMH takes us in the landscape. As previously mentioned, in the best case scenario the hopping can lead to the vicinity of a minimum whose cost function value is smaller than $C(\thv_f,\gamv_f)$. Here SYMH was successful as it allowed to mitigate the effect of noise and improve the quality of the solutions. It might also happen that one lands in the vicinity of a minimum whose cost function value is larger or equal than $C(\thv_f,\gamv_f)$. In this case one simply rejects this particular hop and use Eqs.~\eqref{eq:restriction} to hop to a new set of $\{\widetilde{\thv}_f,\widetilde{\gamv}_f\}$ with different $\vec{\beta}\in\BC$. Finally, if the cost landscape has shifted in such a way that hop does not lead to the vicinity of a minima, one can still perform the second optimization. This would simply correspond to an optimization starting from a new random seed. In all cases, the overhead added by employing SYMH does not change the overall complexity of the optimization.

Note that this high-level description of SYMH is intended as a template than can be employed in many different scenarios. Due to the versatility of the method we do not intend to present here an all-descriptive way of employing SYMH but rather to introduce it as a general method which can be coupled to other optimization and error mitigation techniques. In what follows we present different SYMH-based optimization techniques.

\subsection{Parameter sweeping method}

One of the main challenges that can arise when employing SYMH is the exponentially large number of possible hops one can take (arising from the $2^M$ bitstrings in $\BC$). In this section we present a technique called the \textit{sweeping method} in which instead of exploring all hops, one instead simply explores a reduced sub-set of hops which adds an overhead at most in $\OC(M)$. 

In the sweeping method one starts with $\{\thv_f,\gamv_f\}$ and employs SYMH to obtain a new set $\{\widetilde{\thv}_f,\widetilde{\gamv}_f\}$ such that $\thv_f$ and $\widetilde{\thv}_f$ differ only in the first parameter. That is, the first parameter in the sets $\thv_f$ and  $\widetilde{\thv}_f$ are related according to~\eqref{eq:restriction}. This guarantees that all the parametrized gates in $V(\thv)$ except for the first one remain the same. Then, as previously described one performs a second optimization to find the cost function value at $\{\widetilde{\thv}_f,\widetilde{\gamv}_f\}$, which determines if the shift is accepted or rejected. This procedure is sequentially repeated by sweeping $n_s$ times through all $M$ parameters. We refer the reader to Algorithm~\ref{alg1} for a more detailed description. 
One of the main advantages of this method is that it allows us to identify the parameters that, when shifted, yield the biggest improvement in terms of cost function minimization.

\begin{algorithm}[t ]
\DontPrintSemicolon
\KwIn{Buffered PQC with parameters $\vec{\alpha}=(\thv,\gamv)$ and cost function $C(\vec{\alpha})$; local optimizer \texttt{opt} returning the parameters of a local minimum given a starting point; parameter symmetry \texttt{symh} with input a list of indices of parameters to be shifted; number of sweeps $n_{s}$.}
\KwOut{Converged $C_{f}$ and corresponding parameters $\vec{\alpha}_{f}$.}
    \kwInit{Choose $\vec{\alpha}_0$ at random;
    $\vec{\alpha}_{f} \gets \vec{0}$;
    $C_{f} \gets 0$;
    $\texttt{pulse\_lst} \gets \{\}$.
    }
    Optimize with \texttt{opt} with initial value $\vec{\alpha}_0$ and store result in $\vec{\alpha}_{f}$; $C_{f} \gets C(\vec{\alpha}_{f})$.\;
    \Repeat{\normalfont{no more improvement in $C_{f}$; at most $n_s$ times.}}{
        \For{\normalfont{each parameter in circuit not in $\texttt{pulse\_lst}$ or in buffer}}{
            Use \texttt{symh} with input $\vec{\alpha}_{f}$ to shift the corresponding parameter obtaining $\widetilde{\vec{\alpha}}$;\;
            (Optional) Use \texttt{opt} with input $\widetilde{\vec{\alpha}}$ to find $\widetilde{\vec{\alpha}}_{f}$;\;
            Evaluate $C(\widetilde{\vec{\alpha}}_{f})$.\;
        }
        At end of cycle append the best parameter index to $\texttt{pulse\_lst}$; $\vec{\alpha}_{f}\gets \widetilde{\vec{\alpha}}_{f}$; $C_{f} \gets C(\widetilde{\vec{\alpha}}_{f})$.
    }
    
    (Optional) Do final round of optimization with \texttt{opt}.\;
    \Return{$C_{f}$, $\vec{\alpha}_{f}$.}
    \caption{\sc $\sigma$-Pulse-based\newline parameter sweeping optimization method}
    \label{alg1}
\end{algorithm}

\subsection{Using SYMH for ansatz symmetry breaking and landscape exploration}

Let us now discuss how to implement SYMH to improve the solution quality in problems where the ansatz encodes some additional symmetry beyond that of the $\sigma$-Pulses. In general those symmetries are translate into constraints in the parameters of $V(\thv)$. Specifically,  we analyze the possibility of using SYMH to break those additional constraints and improve the solution quality via the parameter hops obtained through the $\sigma$-Pulse method. 

Unstructured ansatzes such as the hardware efficient ansatz~\cite{kandala2017hardware} have been widely implemented in the literature for problems in which one has little to no information about the solution of the problem. However, there are many tasks in which one possesses knowledge which can be employ to construct the so-called physically-inspired ansatzes. Such information can come in the form of a specific symmetry that the ansatz must preserve~\cite{gard2020efficient}, an adiabatic transformation that must be followed~\cite{qaoa2014,nasaQAOA2019} (as in QAOA), or the operators that the ansatz must contain~\cite{cao2019quantum,bartlett2007coupled,lee2018generalized} (as in UCC). Preserving these additional symmetries during the parameter training can guarantee that the ansatz explores a sub-space of states related to the solution of the problem. 

In most cases, these additional symmetries are translated into parameter constraints which the circuit description of $V(\thv)$ must obey. For instance, in QAOA all the parameters in a given mixing of driving layer are correlated~\cite{streif2020quantum}. Note that in the SYMH formalism this corresponds to exploring certain sub-spaces of $\BC$ where the bistrsings $\vec{\beta}$ respect the ansatz symmetry. 

In the presence of noise, however, it could happen that one can obtain a higher quality solution by starting from the set of ansatz constraint-preserving parameters $\{\thv_f,\gamv_f\}$, and using SYMH to hop to an ansatz constraint-breaking set of parameters. In this case, SYMH allows us to explore circuit configurations inaccessible to the original circuit structure.

\subsection{The Quantum Alternating Operator Ansatz}

We explicitly characterize the symmetries possessed by a Quantum Alternating Operator Ansatz (QAOA)~\cite{nasaQAOA2019}, without a buffer layer. To recap, the QAOA, which generalises the ansatz in the Quantum Approximate Optimization Algorithm~\cite{qaoa2014}, has a layered structure, each layer being composed of two unitaries: a problem unitary $U_P(\beta_i) = e^{-i\beta_i H_P}$, where $H_P$ is the problem Hamiltonian (consisting of $Z$-terms only) and a mixing unitary $U_M(\gamma_i) = e^{-i\gamma_i H_M}$, where typically $H_M = \sum_j X_j$. The parameters of QAOA can therefore be represented as two vectors $\{\vec{\beta}, \vec{\gamma}\}$ each corresponding to each type of unitary. Now let us consider the effect of shifting one of these parameters by $\pi$.

If the parameter belongs to a mixing unitary, the shift will be cancelled by the creation of a $\sigma$-Pulse in the $X$ direction on all the qubits. It can be shown that the commutation of any number of $X$ pulse through a gate consisting of $Z$ terms of any order generates no additional pulses, but leads to the parameter of the gate acquiring a negative sign. Therefore, the $X$ pulses generated by such a shift can only be annihilated by a restoring shift of another mixing parameter. Therefore, a symmetry is the following:
\begin{align}
    &\gamma_i \rightarrow \gamma_i + \pi \nonumber\\
    &\gamma_j \rightarrow \gamma_j + \pi, \, i < j \nonumber\\
    &\beta_k \rightarrow -\beta_k, \, \forall i < k \leq j \nonumber
\end{align}

If instead we focus on the problem unitary, we see that the parameter shift will create, on a given qubit, one $Z$ pulse for each $Z$ acting on that qubit. Since the mixing unitary presents one $X$ rotation for each qubit, one concludes that, in order to be able to commute all of the pulses past a mixing unitary with only a sign modification of its parameter, there must be the same number of $Z$ pulses on all the qubits. Therefore, the ansatz will only possess such a symmetry if the problem Hamiltonian features the same number of $Z$s on each qubit. This happens, for example, in MaxCut problems on n-regular graphs. In this specific case, when n is odd each shift will generate one total $Z$ per qubit, and for any $i,\, j$ with $i < j$ the symmetries are:
\begin{align}
    &\beta_i \rightarrow \beta_i + \pi, \, \beta_j \rightarrow \beta_j + \pi, \,  \nonumber\\
    &\gamma_k \rightarrow -\gamma_k \; \forall k \, i < k \leq j . \nonumber
\end{align}
Otherwise, for n even all the pulses will cancel and the symmetries will simply be:
\begin{equation}
    \beta_i \rightarrow \beta_i + \pi \; \forall i .\nonumber
\end{equation}
Identifying these symmetries is significant, because they imply, if not a complete parameter space reduction, that restricting the range of some parameters will not affect the result of the algorithm.

\section{Implementations}\label{sec:numerics}

In this section we present heuristic results where we employ SYHM to improve the solution quality of VQAs in the presence of quantum noise. Specifically, we simulate VQAs for two implementations: variational quantum compiling, and for a Variational Quantum Eigensolver problem. 

\subsection{Variational Quantum Compiling}\label{sec:VQC}

Quantum compiling~\cite{chong2017programming,haner2018software,venturelli2018compiling} refers to the task of transforming a high-level algorithm into a low-level code that a quantum hardware can efficiently implement. In the near-term, one of the main applications for quantum compiling is to transforming a unitary with a deep quantum circuit description into a shorter depth gate sequence which mitigates the effect of noise. Several Variational Quantum Compiling (VQC)~\cite{jones2018quantum, heya2018variational,Khatri2019quantumassisted} architectures have been recently introduced where one trains the parameters in a short-depth PQC so that its outputs approximate those of a target unitary $U$. 

Here we consider a quantum compiling application where the goal is  to train  the parameters in the PQC so that  $V_B(\thv,\gamv)\ket{\vec{0}}=U\ket{\vec{0}}$, where $U$ is the $W$-state preparation circuit for three qubits (see~\cite{sharma2019noise} for an explicit circuit), and where $\ket{\vec{0}}=\ket{0}^{\otimes 3}$ is the all-zero state. Explicitly, $U\ket{\vec{0}}=\ket{W}$, where $\ket{W}$ is the three qubit $W$-state. For simplicity, we consider the cost function 
\begin{equation}
    C = 1 - \Tr\left[U\dya{\vec{0}}U\ad V_B(\thv,\gamv) \dya{\vec{0}}V_B\ad(\thv,\gamv)\right] \,,
\end{equation}
which vanishes if $V_B(\thv,\gamv)\ket{\vec{0}} =\ket{W}$ (up to a global phase). 

\begin{figure}[t]
    \centering
    \includegraphics[width=0.45\textwidth]{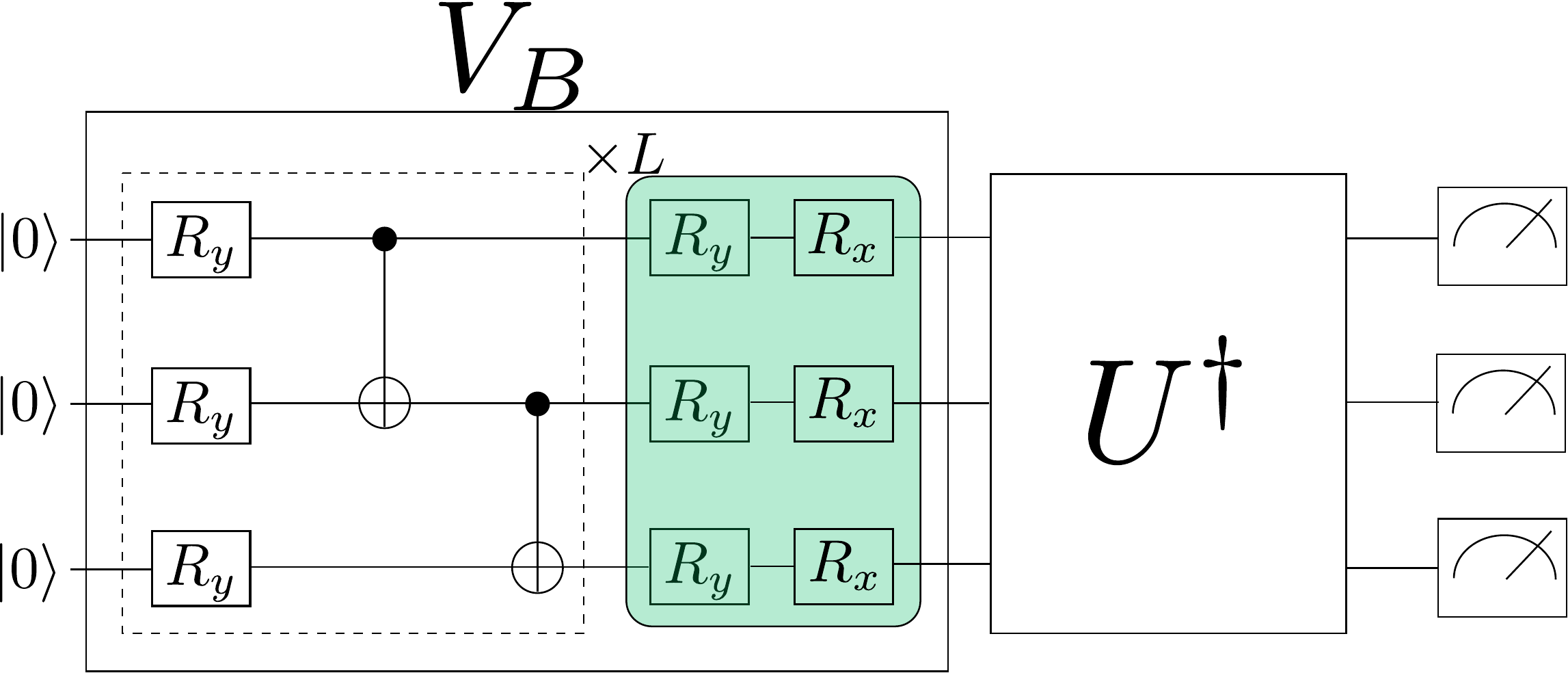}
    \caption{Circuit employed for the $W$-state quantum compilation implementation. Here we have indicated with a box the ansatz for $V_B(\thv,\gamv)$. By measuring the probability of all qubits being on zero $P(\vec{0})$, the cost can be computed as $C=1-P(\vec{0})$. Here  the buffer layer is only composed of rotations about the $y$ axis, as those are sufficient.}
    \label{fig:VCansatz}
\end{figure}

As outlined in Fig.~\ref{fig:VCansatz}, for the buffered PQC, we choose a layered hardware-efficient ansatz of the form
\begin{equation}
    U(\thv, \vec{\gamma}) = U_B(\vec{\gamma}) \left(\prod_{i=1}^L U_i(\thv_i) \right)\,,
\end{equation}
where each $U_i(\thv_i)$ consists of single qubit rotations followed by CNOTs. Note that here $U_B(\vec{\gamma})$ is only composed of $R_y$ rotations as we can only create $\sigma_y$ pulses.

\begin{figure}[t]
    \centering
    \includegraphics[width=0.5\textwidth]{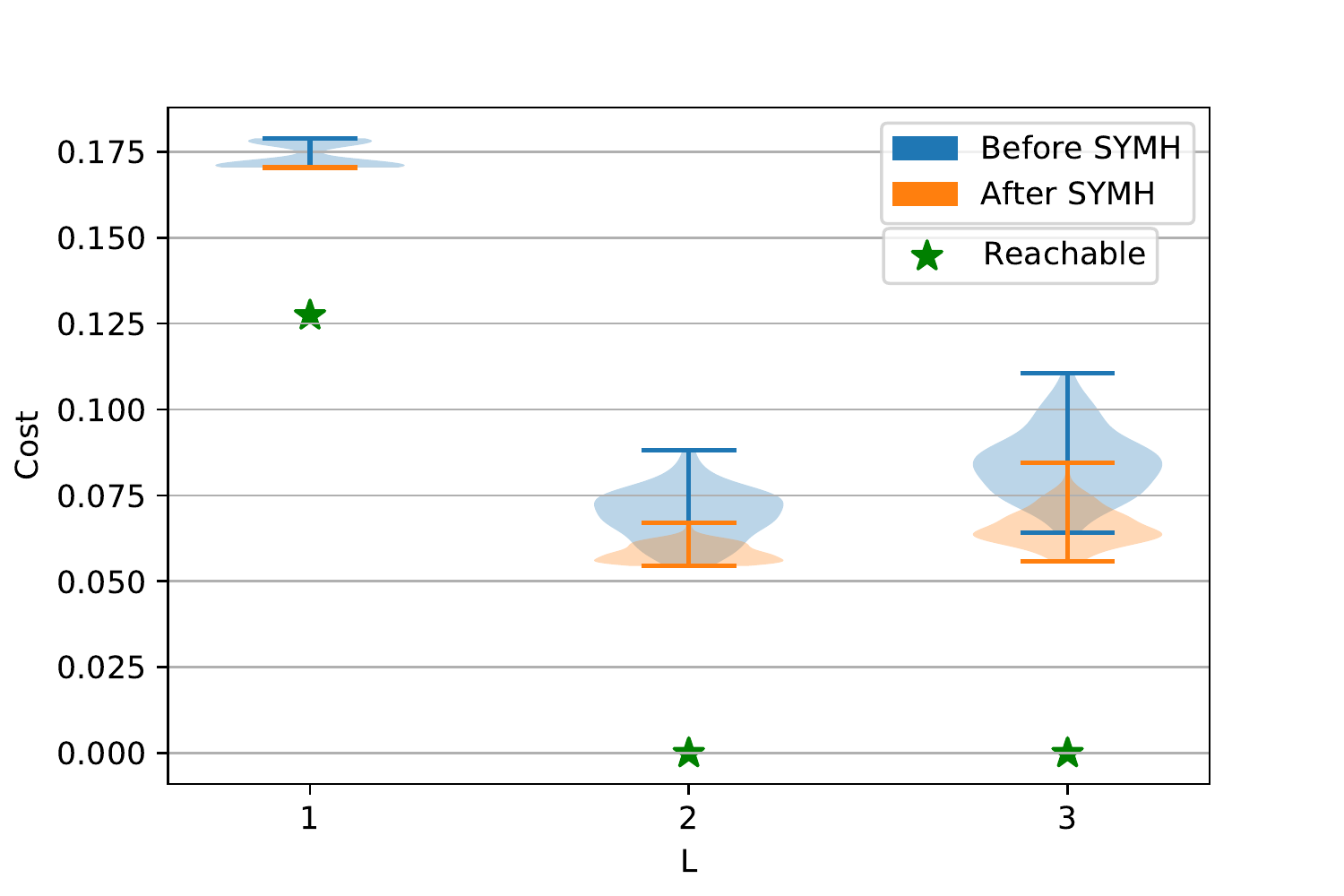}
    \caption{Numerical results for the $W$-state quantum compiling implementation. We show cost function value versus the number of layers $L$ in $V(\thv)$ before and after implementing SYMH, for a circuit subject to a realistic noise model with amplitude damping noise. The stars indicate the cost function value reachable in the absence of noise. For each $L$, we show the spread of the cost function value results for $100$ randomly initialized instances of the optimization, with the horizontal lines indicating the best and worst run.}
    \label{fig:VCresults}
\end{figure}

In Fig.~\ref{fig:VCresults} we present our numerical results. Here we employed a noisy quantum circuit simulator with realistic amplitude damping noise acting after every gate (including idle gates) obtained from the average T1 and T2 parameters and gate times of the \texttt{ibmq\_melbourne} quantum computer. For gradient descent we used the COBYLA~\cite{PowellCOBYLA} optimizer, initialized at random angles. In addition, we simulated the circuit with an ansatz composed of $L=1,2,3$ layers. In a noiseless scenario, a single layer is not enough to prepare the $W$ state, while $L=2,3$ can reproduce the desired target state. For each number of layers we ran $100$ randomly initialized simulations, and once the first optimization was completed we implemented the parameter sweeping method of  Algorithm~\ref{alg1} with $n_s=4$ maximum sweeps. 

As seen in Fig.~\ref{fig:VCresults}, for all values of $L$ employing SYMH leads to a systematic improvement in the cost function value. This improvement can be measured by taking either the best run (out of $100$) or by taking the average of all runs before and after SYMH. In addition we can see that the improvement increases with $L$ as circuits with a longer depths accumulate more noise, hence leading to a larger possible improvement. 

\subsection{Variational Quantum Eigensolver}

In this section we present our results for a Variational Quantum Eigensolver (VQE) implementation. Here, the goal is to train a PQC such that $V_B(\thv,\gamv)$ prepares the ground-state of a given Hamiltonian $H$. Specifically, we consider $H$ to be the $SU(2)$-symmetric Heisenberg $XXX$ model on $n$ qubits
\begin{align}
    H_{\text{XXX}} = &\sum^{n}_{i=1} \sigma^i_x \sigma^{i+1}_x + \sigma^i_y \sigma^{i+1}_y + \sigma^i_z \sigma^{i+1}_z\,.
\end{align}
Where we assume period boundary conditions so that $n+1\equiv 1$. Here the cost function is simply given by
\begin{equation}
    C=\Tr[HV(\thv)\dya{\psi}V\ad(\thv)]\,,
\end{equation}
where $\ket{\psi}$ is an efficiently preparable input state.

\begin{figure}[t]
    \includegraphics[width=0.48\textwidth]{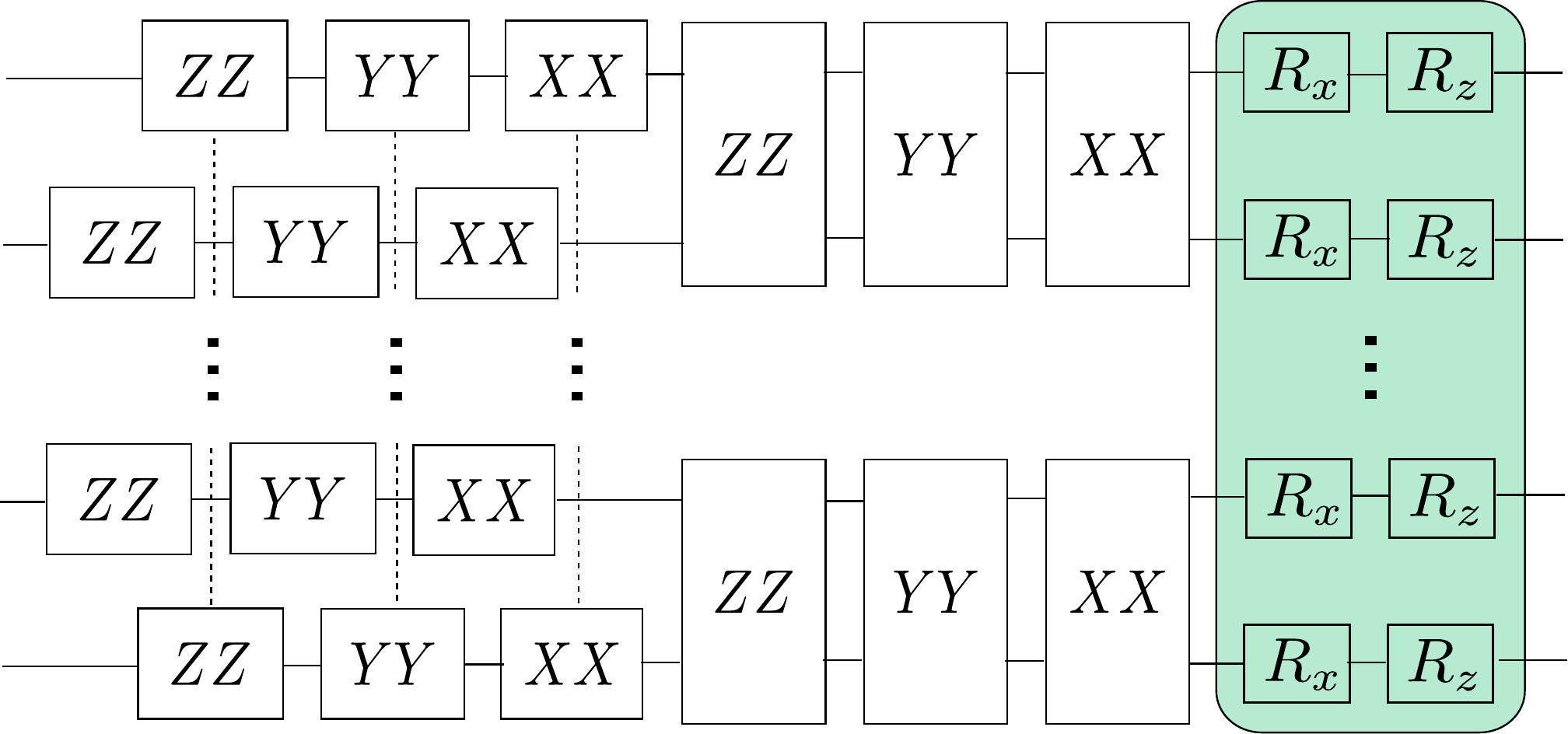}
    \caption{Single layer of the Hamiltonian Variational Ansatz for the $XXX$ model. Here, the gates $XX$ indicate a unitary of the form $e^{- i \theta X_iX_{i+1}}$. We employ a similar notation for $YY$ and $ZZ$. The angles in the buffer layer are set to zero during the initial optimization to preserve the ansatz constraints.}
    \label{fig:QAOAnsatz}
\end{figure}

For the PQC ansatz we consider a subclass of the QAOA called Hamiltonian Variational Ansatz (HVA), first introduced in~\cite{wecker2018towards}. Specifically,  we follow the circuit structure for HVAs of~\cite{wiersema2020exploring}, where we split  $H_{\text{XXX}}$ into two summations with the $i$ index being even and odd. That is,  $H_{\text{XXX}}=H_{\text{odd}}+H_{\text{even}}$ so that the ansatz can be expressed as
\begin{equation}
    V(\thv) = \prod_{i=1}^L e^{-i\theta_i H_{\text{even}}/2} e^{-i\theta_i H_{\text{odd}}/2} \,.
\end{equation}
In addition, here we choose the initial state to be a tensor product of the Bell state $\ket{\psi} = \frac{1}{2^{n/4}} \left(\ket{01} - \ket{10}\right)^{\otimes n/2}$, which is the ground state of $H_{\text{even}}$. As shown in Fig.~\ref{fig:QAOAnsatz},  the circuit description of  $V(\thv)$ can be obtained from a first order Trotter expansion of the $e^{-i\theta_i H_{\text{even}}/2}$ and $e^{-i\theta_i H_{\text{even}}/2}$ unitaries and hence  features alternating layers of $XX$, $YY$ and $ZZ$ interactions, first on odd qubits and then on even qubits.

\begin{figure}[!t]
    \centering
    \includegraphics[width=0.75\linewidth]{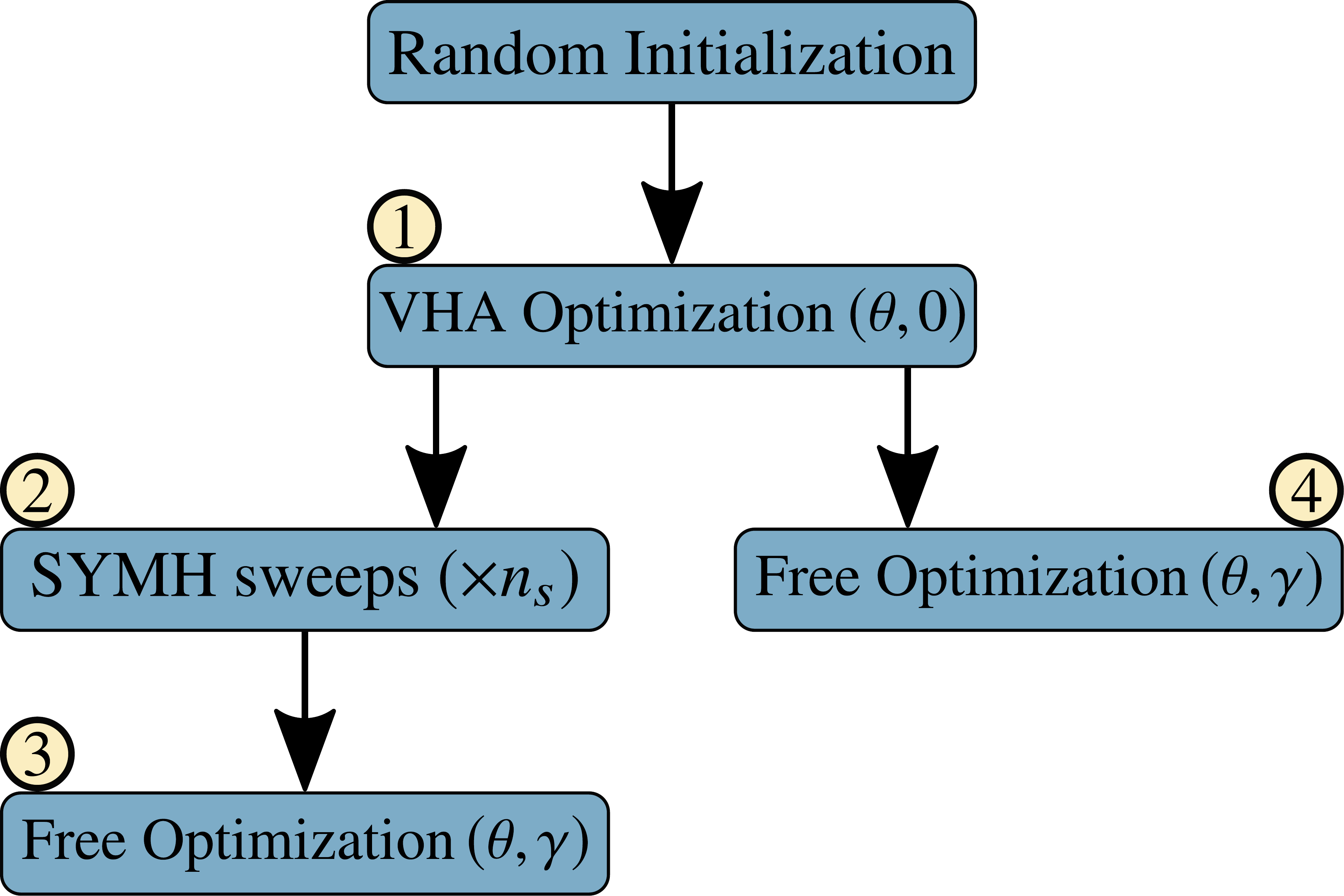}
    \caption{Flow diagram of the optimization schedules employed in the VQE implementation. (1) The parameters in the VHA are optimized respecting the ansatz constrains. (2) We employ SYMH to hop in the landscape. (3)-(4) In a free optimization, the parameters $\thv$ and $\gamv$ are independently optimized, hence breaking the constrains in the HVA.  }
    \label{fig:Schedules}
\end{figure}

Let us remark that the parameters in $V(\thv)$ encode additional problem symmetries  as all the gates in a layer are constrained to being identical. To show that SYMH can be used to break these constraints and improve the solution quality in the presence of noise, our heuristics where perform with the optimization schedules of Fig.~\ref{fig:Schedules}. First, we set the parameters in the buffer layer to be $\gamv=\vec{0}$ so that $U_B(\thv)=\id$. Then, we randomly initialize the $\thv$ and optimize them respecting their correlation (see Fig.~\ref{fig:Schedules}(1)). As indicated in Fig.~\ref{fig:Schedules}(2), we then implement SYMH to break the VHA constraints, and hop in the landscape. This hopping can be followed by a second optimization, which we call a ``free'' optimization, where the VHA constraints are broken and all parameters are independently optimized (see Fig.~\ref{fig:Schedules}(3)). To verify that the improvements arise from the SYMH and not from the free optimization, for each run we additionally perform a free optimization without SYMH (Fig.~\ref{fig:Schedules}(4)). 

\begin{figure}[t]
    \centering
    \includegraphics[width=0.48\textwidth]{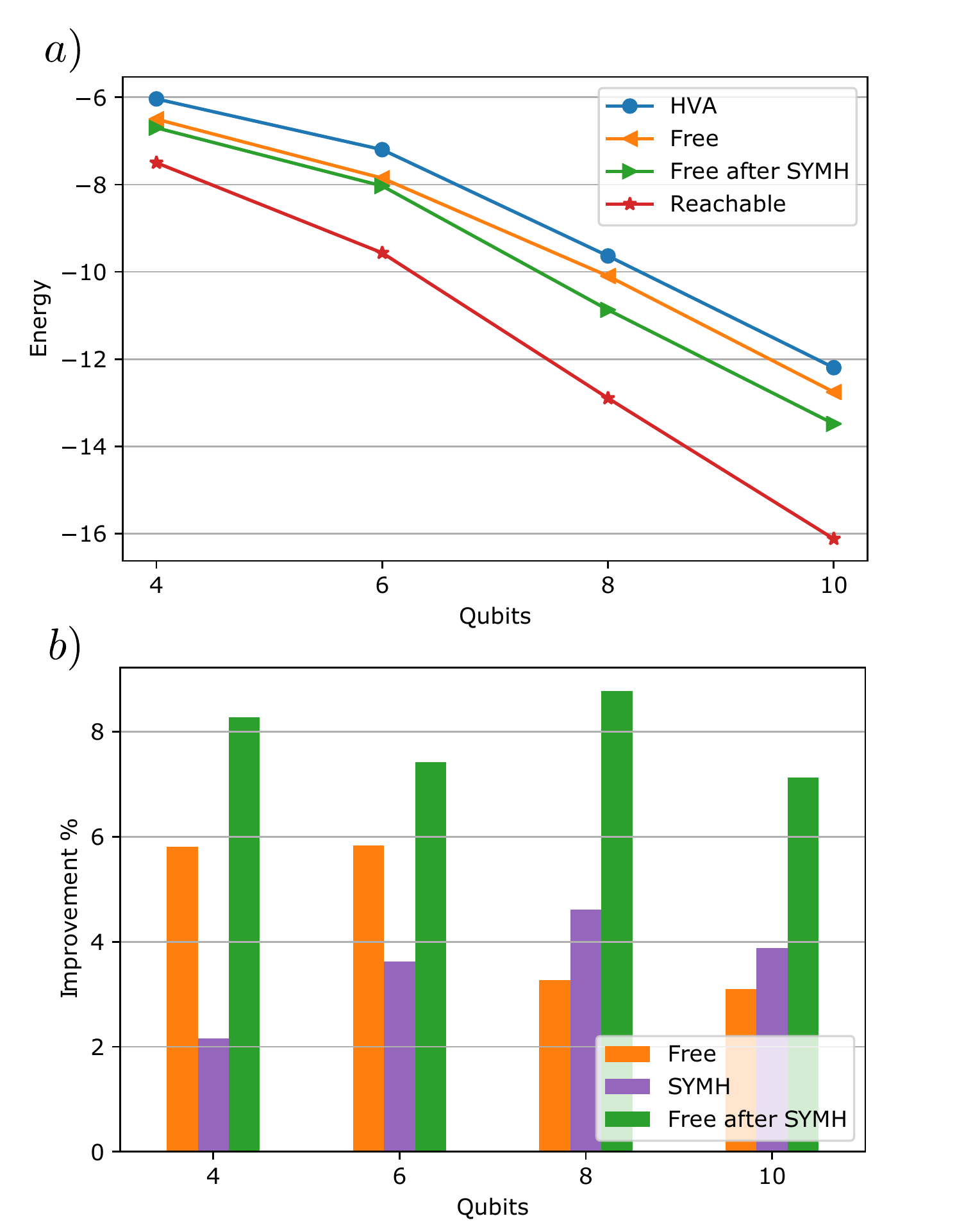}
    \caption{Numerical results for the Variational Quantum Eigensolver implementation. (a) Cost function value (for the best run) versus number of qubits $n$. Each curve represents a different optimization scheme as shown in Fig.~\ref{fig:Schedules}. For all values of $n$ employing SYMH and breaking the ansatz parameter constrains leads to the best improvement.  (b) Cost function value improvement versus number of qubits for different optimization schemes. The improvement is defined in Eq.~\eqref{eq:improvement}. For all values of $n$ the largest improvement arises by employing SYMH and breaking the ansatz parameter constrains.  }
    \label{fig:SymmetryQAOAresults}
\end{figure}

The results from our numerical simulations are presented in  Fig.~\ref{fig:SymmetryQAOAresults}. Here we considered VQE problems with $n=4,6,8,10$ qubits and where the ansatz is composed of a single layer. Moreover, we employ the same noise model and optimizer as the one described in the previous section.  For each $n$ we ran 100 instances of the optimization. In Fig.~\ref{fig:SymmetryQAOAresults}(a) we present the best run for each number of qubits and for the different optimization method in Fig.~\ref{fig:Schedules}. As shown, one can always improve the solution quality by breaking the ansatz constrains. However, the best solution is always achieved when employing the SYMH. Meaning that the optimal improvement follows from hoping in the landscape.  

A more detailed comparison of the cost function improvement  for the different methods is presented in Fig.~\ref{fig:SymmetryQAOAresults}(b), where the improvement is defined as 
\begin{equation}\label{eq:improvement}
    \text{improvement \%}=100 \times \frac{E_{f} - E_{HVA}}{E_{GS}}\,.
\end{equation}
Here, $E_{GS}$ denotes the true ground state energy $E_{HVA}$ is the energy obtained from $(1)$ in Fig.~\ref{fig:Schedules} and $E_{f}$ is the final energy from the optimization schemes $(2)$, $(3)$ and $(4)$ of Fig.~\ref{fig:Schedules}. As shown in Fig.~\ref{fig:SymmetryQAOAresults}(b), employing SYMH and breaking the parameter constrains always seems to leads to the best improvement. In fact, for all values of $n$ the improvement is larger than $7\%$.

\section{Discussion}

Analyzing the cost function landscape of variational quantum algorithms and quantum neural networks is a fundamental task to improve their performance. While some rigorous results have been derived that examine the connection between the ansatz and the cost landscape, much still remains to be done. 
In this work we discussed two phenomena related to the cost function landscape, the first phenomenon being an exponentially large symmetry in the parameters of a PQC, and the second pertains to how quantum noise affects these symmetries. The latter provides the theoretical grounding for the SYMH optimization method. We will now separately summarise these results and discuss their implications.

\subsection{Exponential symmetry in PQCs}

The first phenomenon can be condensed into the following: for buffered PQCs, there are exponentially many sets of symmetric parameters $\thv$ and $\thv'$ such that $V(\thv)=V(\thv')$ (up to a global phase). These symmetries translate in turn into exponential degenerancies in the cost landscape. To understand and analyze these symmetries we have introduced the so-called $\sigma$-Pulse method. The main idea behind this method is the creation, propagation, and absorption of virtual gates in the PQC, which allow us to obtain symmetric sets of parameters. Despite the simple interpretation of the $\sigma$-Pulse method, its implications are non-trivial. For instance, we can show that all the relevant features of the cost landscape can be found in a subspace of the parameter hyperspace, hence providing an exponential reduction of the search space of variational quantum algorithms.

To the best of our knowledge, the current work represents the first discussion of such a broad degeneracy in quantum parameterized quantum circuits (altho subsequent works have studied existence of exponentially many local minima of overparameterized quantum neural networks~\cite{you2021exponentially}).
Nonetheless, there exists a connection with concepts in Measurement-Based Quantum Computing (MBQC) \cite{briegel2009measurement}, and with the concept of flow \cite{danos2006determinism} in particular. The symmetries presented here are related to the correction operations following a measurement to a qubit. A variant of the "angles can be restricted to $[0, \pi)$" principle is even known for such measurement operations~\footnote{Private communication with Will Simmons}.

In the context of variational algorithms, the result is however novel and may have numerous implications.
For instance, the fact that the noiseless quantum landscape has been demonstrated to consist of an exponentially large number of translated copies of a single landscape ``unit cell'' means that the informational content of the landscape is actually greatly reduced. As such, one may hope to construct quantum algorithms that are more sample-efficient by exploiting these regularities. Beyond variational algorithms, quantum compilers might leverage the symmetries to reduce the single-qubit gate count from static circuits. Indeed, any $R_{\mu}(\pi)$ gate may be eliminated by shifting it to an identity gate via an appropriate $\sigma$-Pulse. This might also be done in an approximate manner, for example if the rotation is only a small $\epsilon$ away from $\pi$. However such an approach would leave the circuit's $T$ count invariant.

\subsection{Noise-broken symmetries and SYMH}

The second theoretical contribution considers what happens when noise in introduced in such a buffered PQC. First we rigorously showed that the parameter symmetries are preserved under the action of unital Pauli noise, implying that dephasing and depolarizing noise acting throughout the PQC have no effect on the overall symmetric structure of the cost landscape. We then proved that non-unital Pauli can break the parameter symmetries and hence the degenerancies in the cost landscape. This result implies that, when training in the presence of noise, some of the previously exponentially degenerate global minima can become local minima. Hence, optimization strategies that randomly initialize the parameters could converge to one of those local minima and not obtain an optimal solution. 

To mitigate the effect of noise when training in degeneracy-broken landscapes, we introduced a novel optimization method which we call the Symmetry-based Minima Hopping (SYMH) optimizer. SYMH employs the parameter symmetries to hop around the landscape and attempt to converge to more noise resilient minima. The main advantage of SYMH is its versatility, in the sense that it can be easily combined with any optimization or error mitigation technique without significantly increasing the computational overhead.

To showcase the effectiveness of SYMH we numerically simulated two variational quantum algorithms in the presence of quantum noise. Namely, we implemented a quantum compiling task and the Variational Quantum Eigensolver. In both cases we heuristically showed that employing SYMH when optimizing consistently improves the solution quality. We remark that these are preliminary results, and more thorough approaches to constructing optimizers with SYMH might well prove to be even more fruitful. Hence, SYMH is an additional tool in the quantum variational toolbox, and it can be regarded as a quantum-aware optimizer that accounts for symmetries and symmetry breaking in noisy PQCs. 

At the same time, one should be mindful that SYMH in its present formulation cannot be employed as a standalone optimizer, but must be paired with a suitable local optimization routine. The local gradient descent step can be kept minimal, but it is nonetheless unavoidable as without it the optimization would be constrained to a discrete grid of points connected by $\sigma$-Pulse symmetries, where it is highly unlikely that the global minimum lies.

To better gauge the utility of SYMH, it is helpful to compare it with similar but non-quantum native approaches to global optimization. The classical machine learning literature offers several such algorithms, the main class being simulated annealing (SA) optimizers~\cite{kirkpatrick1983optimization}, which include Metropolis-based algorithms like the particle-collision algorithm~\cite{sacco2005new}. These explore a number of random candidate solutions and select the best ones based on a probabilistic method. The similarity with SYMH is that such algorithms are designed to evade local minima and explore as much of the landscape as possible. Some of them may also incorporate a gradient-based local subroutine~\cite{rios2010metropolis}. However, one considerable difference is that these algorithms are still local, as at each step they make gradual deviations away from the starting point. Setting a large step size would lead to unpredictable variations in the cost. Instead, SYMH is in principle able to make large jumps across the landscape while limiting the change in cost, as the algorithm is rooted on knowledge of the underlying symmetries of the landscape. Similar considerations also hold for the method of gradient descent with warm restarts~\cite{loshchilov2016sgdr}. Nonetheless, it is an interesting question for future research if a combination of SA-style approaches and SYMH jumps would be more effective than either for a quantum problem.

Interestingly, the SYMH method also has a natural connection with the broad family of randomized methods for noise mitigation, which include randomized compiling and benchmarking~\cite{kern2005quantum,wallman2016noise}, twirling~\cite{moussa2012practical} and probabilistic error cancellation~\cite{temme2017error}. 
In fact the parameter symmetries obtained through our $\sigma$-Pulse method are a special subset of twirling operations that preserve the structure of the ansatz. Nonetheless, the aforementioned methods involve a modification of the circuit via the addition of extra Pauli gates, while we achieve this implicitly by shifting the parameter values. Another difference is that the symmetries are specific to variational circuits, and violate the assumption of Clifford operations required by most twirling-based methods.
As noise mitigation techniques, SYMH and randomized error mitigation methods are distinct. The latter aim at converting general noise channels into Pauli ones by a symmetrization procedure, while SYMH leverages the broken symmetries to achieve an improved cost function. In addition SYMH does not require any averaging operation, leading to a smaller overhead.

Despite its merits, SYMH presents some limitations that we now discuss. First, the parameter sweeping algorithm is purely heuristic and comes with no guarantee of convergence. Certainly, if one were given knowledge of the precise noise channels affecting the system, it is reasonable to expect that more refined optimization methods could be identified. As such, we expect that more advanced versions of SYMH can be derived given knowledge of the noise structure.  Still, the numerical results found here suggest that the sweeping method is able to improve the cost function significantly, see Figure~\ref{fig:SymmetryQAOAresults}.

Secondly, even with an ideal optimization schedule SYMH would only be effective up to unital channels, which cannot be corrected by SYMH. These channels, e.g. Pauli channels, are widely present in quantum computers~\cite{flammia2020efficient}. However, is it also true that many noise mitigation methods exist that are effective for unital Pauli channels~\cite{temme2017error, li2017efficient, endo2018practical, urbanek2021mitigating}, and such methods may be easily integrated alongside SYMH. Ultimately we propose SYMH as a first step towards leveraging the quantum landscape structure to optimize cost functions. Moreover, SYMH should not be considered as a standalone technique, but one to be used in conjunction with more standard strategies.

Several future research directions follow for the SYMH method. First, we heuristically verified that hopping does not lead to a critical point, but rather to the valley of a minimum.  Analyzing how much the minima can shift could provide additional guarantees for SYMH without a second optimization. Second, in our numerics we observed that shifting some parameters leads to the greatest improvements, but no clear pattern emerged. We leave for future work the analysis of this phenomenon.

\section{ACKNOWLEDGEMENTS}

We thank Kunal Sharma and Lukasz Cincio for helpful discussions. We acknowledge the help of Alexander Cowtan and Ross Duncan through discussions regarding ZX-calculus and its applications to circuit compilation. We are grateful to Will Simmons for pointing out the connection between the parameter symmetries and the concept of flow in MBQC. EF was initially supported by the U.S. Department of Energy (DOE) through a quantum computing program sponsored by the LANL Information Science \& Technology Institute. EF and IR acknowledge the support of the UK government department for Business, Energy and Industrial Strategy through the UK national quantum technologies programme. EF acknowledges the support of an industrial CASE (iCASE) studentship, funded by the UK Engineering and Physical Sciences Research Council (grant EP/T517665/1). MC acknowledges initial support from the Center for Nonlinear Studies at LANL. AA was initially supported by the Laboratory Directed Research and Development (LDRD) program of LANL under project number 20190065DR. PJC acknowledges initial support from the LANL ASC Beyond Moore's Law project. This work was supported by the U.S. DOE, Office of Science, Office of Advanced Scientific Computing Research, under the Quantum Computing Application Teams~program.

\bibliographystyle{unsrtnat}
\bibliography{ref.bib}

\clearpage



\appendix

\vspace{0.5in}

\begin{center}
  { \bf \MakeUppercase{Appendix}}
\end{center}


In this appendix we provide of our main results, with Appendix~\ref{app:1} containing  the proof of Proposition~\ref{prop1} and Corollary~\ref{cor:reduction}. Then, in  Appendix~\ref{app:2} we present  the proofs of Theorem~\ref{theo:1} and Theorem~\ref{theo:2}.

\section{Proof of Proposition~\ref{prop1} and Corollary~\ref{cor:reduction}}~\label{app:1}

In what follows we present the proof for Proposition~\ref{prop1}.
\begin{proof}
Let us first recall that, as mentioned in the main text, there are $M$ parameters  in the PQC $V(\thv)$. Moreover let us denote as $\GC$ the set of all possible generator choices. The number of distinct symmetric sets of parameters $\{\thv,\gamv\}$ obtained trough the $\sigma$-Pulse method is given by  the cardinality of $\GC$:
\begin{equation}\label{eq:numparam}
    \vert\GC\vert=\sum_{\alpha=0}^{M}\binom{M}{\alpha}=2^{M}\,.
\end{equation}
Similarly, we can count the number of symmetric sets of parameters as the number of bitstrings $\vec{\beta}$ in $\BC$ of length $M$, which is precisely $2^M$.
\end{proof}

Now we prove Corollary~\ref{cor:reduction}. In particular we show that given any $\{\thv,\gamv\}$ one can always find a set $\{\widetilde{\thv},\widetilde{\gamv}\}$ with $\vec{\beta}=\vec{0}$.

\begin{proof}

Let us assume that the parameters in $\thv=(\theta_1,\theta_2,\ldots)$ are order by layer,  where a layer consists of quantum gates that can be performed in parallel and where the first layer contains the first gates in $V(\thv)$. In general we can assume without loss of generality that the angles in $\thv$ are in $[0,2\pi)$. We now describe a sequential procedure that can be used  to obtain the vector $\widetilde{\thv}$ where every parameter $\widetilde{\theta_j}\in \widetilde{\thv}$  not in the buffer layer are in the reduced domain   $[0,\pi)$. 

If $\theta_1\in[0,\pi)$ then we do nothing, but if  $\theta_1\in[\pi,2\pi)$ we create and forward propagate a $\sigma$-Pulse. According to Eq.~\eqref{eq:pipulse}, this will add $\pi$ to $\theta_1$, which maps it to the interval $[0,\pi)$. This procedure is then sequentially repeated for each parameter in $\thv$ not in the buffer layer. We remark that since $\sigma$-Pulse propagate forward in the circuit, creating a pulse in $\theta_j$ does not affect any angle $\theta_k$ with $k<j$. Moreover, we know from~\eqref{eq:pipulserot} that as the $\sigma$-Pulses propagate they can add a minus sign to other angles in $\thv$.  Hence, if a given $\theta_j$ that was originally in $[\pi,2\pi)$ picked up a minus sign then we do nothing as it will now be in $[0,\pi)$. On the other hand, if it was in $[0,\pi)$  we have to create a $\sigma$-Pulse to map it to $[0,\pi)$. Note that at the end of this procedure every parameter not in the buffer layer will be mapped to the reduced domain $[0,\pi)$.
\end{proof}

\pagebreak

\section{Proof of our main theorems}~\label{app:2}

\subsection{Proof of Theorem~\ref{theo:1}}

Let us start by presenting a more detailed definition of a unital Pauli channel:

\noindent \textbf{Unital Pauli channels}. 
A Pauli noise channel corresponds to the action of random Pauli operators according to a given probability distribution. Let $\PC_U$ denote an $n$-qubit Pauli channel. The action of $\PC_U$  on any given $n$-qubit Pauli operator is given by  
\begin{equation}\label{eq:Pauli-channel}
\PC_U(X^{\vec{a}}Z^{\vec{Z}})=\sum_{\vec{l},\vec{k}}p_{\vec{l},\vec{k}}X^{\vec{l}}Z^{\vec{k}}(X^{\vec{a}}Z^{\vec{b}})(X^{\vec{l}}Z^{\vec{k}})^\dagger,
\end{equation}
where $0\leq p^A_{\vec{l},\vec{k}}\leq 1$, and $\sum_{\vec{l},\vec{k}}p^A_{\vec{l},\vec{k}}=1$. By using the fact that
\begin{equation}\label{eq:commute-Pauli}
    X^{\vec{l}}Z^{\vec{k}}=(-1)^{\vec{l}\cdot\vec{k}}Z^{\vec{k}}X^{\vec{l}}\,,
\end{equation}
we find 
\begin{align}
\PC_U(X^{\vec{a}}Z^{\vec{b}})&=\sum_{\vec{l},\vec{k}}p_{\vec{l},\vec{k}}X^{\vec{l}}Z^{\vec{k}}X^{\vec{a}}Z^{\vec{b}}Z^{\vec{k}}X^{\vec{l}} \nonumber\\
&=\sum_{\vec{l},\vec{k}}(-1)^{\vec{a}\cdot\vec{k}}(-1)^{\vec{b}\cdot\vec{l}} p_{\vec{l},\vec{k}}X^{\vec{a}}Z^{\vec{b}} \nonumber\\
&=p_{\vec{a},\vec{b}}X^{\vec{a}}Z^{\vec{b}},
\end{align}
where $p_{\vec{a},\vec{b}}= \sum_{\vec{l},\vec{k}}(-1)^{\vec{a}\cdot\vec{k}}(-1)^{\vec{b}\cdot\vec{l}} p_{\vec{l},\vec{k}}$ and $-1\leq p_{\vec{a},\vec{b}}\leq 1$ for all $\vec{a},\vec{b}\in\{0,1\}^n$. 

\bigskip

We now prove Theorem~\ref{theo:1}.

\begin{proof}
Let us now consider a buffered circuit $V_B(\thv,\gamv)$ which is implemented in the presence of a unital Pauli noise mode as presented in Definition~\ref{def:Unital}. Here we show that propagating the pulses (Pauli operators) trough the circuit does not change the unitary being produced as the Pauli operators commute with unital Pauli noise. 

Let us now consider the channel $\VC_B$ which implements the unitary $V_B(\thv,\gamv)$. This channel can be expressed as 
\begin{equation}
    \VC_B=\UC_B\circ\VC_L\circ\cdots\circ\VC_1\,,
\end{equation}
where $\VC_l$ is the channel that implements the unitaries in the $l$-th layer, and where $\UC_B$ the channel that implements the buffer unitary. From the $\sigma$-Pulse method we know that we can find a symmetric set of parameters by creating a sigma pulse, propagating it, and absorbing it in the buffer layer. In the channel notation this procedure can be expressed as
\begin{itemize}
    \item Creation of a primary $\sigma$-Pulse:
    \begin{equation}
        \VC_B=\UC_B\circ\VC_L\circ\cdots\circ\Sigma_1\circ\widetilde{\VC}_1\,,
    \end{equation}
    with $\Sigma_1$ the channel that implements the $\sigma$-Pulse, and with $\widetilde{\VC}_1$ the parameter shifted unitary. 
    \item Propagation of the $\sigma$-Pulses:
        \begin{equation}
        \VC_B=\UC_B\circ\Sigma_L\circ\widetilde{\VC}_L\circ\cdots\circ\widetilde{\VC}_1\,,
    \end{equation}
    where now $\Sigma_L$ is the channel that implements the primary and secondary $\sigma$-Pulses.
    \item Absorption of the  $\sigma$-Pulses:
            \begin{equation}
        \VC_B=\widetilde{\UC_B}\circ\widetilde{\VC}_L\circ\cdots\circ\widetilde{\VC}_1\,.
    \end{equation}
\end{itemize}
Here we can see that the channel $\VC_B$ remains unchanged, meaning that $V_B(\thv,\gamv)=V(\widetilde{\thv},\widetilde{\gamv})$.

Let us now analyze this procedure in the presence of noise. The noisy version of the channel that implements $V_B(\thv,\gamv)$ can be expressed as 
\begin{equation}
    \widehat{\VC_B}=\PC_U^{(L+2)}\circ\UC_B\circ\PC_U^{(L+1)}\circ\VC_L\circ\PC_U^{(L)}\cdots\circ\PC_U^{(1)}\circ\VC_1\circ\PC_U^{(0)}\,,
\end{equation}
with $\PC_U^{(l)}$ the noisy channel acting after every layer of gates. Once a $\sigma$-Pulse has been created we will have 
\begin{equation}
    \widehat{\VC_B}=\PC_U^{(L+2)}\circ\UC_B\circ\PC_U^{(L+1)}\circ\VC_L\circ\PC_U^{(L)}\cdots\circ\PC_U^{(1)}\circ\Sigma_1\circ\widetilde{\VC}_1\circ\PC_U^{(0)}\,.
\end{equation}

As we now show, any unital noisy channels $\PC_U^{(i)}$ always commute with the channels $\Sigma_k$ that implement $\sigma$-Pulses. Explicitly, the action of  on any given $n$-qubit Pauli operator is given by
\begin{align}\label{eq:pulse-channel}
    \Sigma_k(X^{\vec{a}}Z^{\vec{b}})&=X^{\vec{p}}Z^{\vec{q}}X^{\vec{a}}Z^{\vec{b}} (X^{\vec{p}}Z^{\vec{q}})\ad\nonumber\\
    &=(-1)^{\vec{a}\cdot\vec{q}}(-1)^{\vec{b}\cdot\vec{p}}X^{\vec{a}}Z^{\vec{b}}\,.
\end{align}
Hence, using Eqs.~\eqref{eq:commute-Pauli}, \eqref{eq:pulse-channel} and~\eqref{eq:Pauli-channel} we have that the following chain of equalities always hold
\small
\begin{align}
\mathcal{P}_{U}^{(i)}\circ\Sigma_k(X^{\vec{a}}Z^{\vec{b}})&=\sum_{\vec{l},\vec{k}}p_{\vec{l},\vec{k}}X^{\vec{l}}Z^{\vec{k}}X^{\vec{p}}Z^{\vec{q}}(X^{\vec{a}}Z^{\vec{b}})Z^{\vec{q}}X^{\vec{p}}Z^{\vec{k}}X^{\vec{l}}\nonumber\\
&=\sum_{\vec{l},\vec{k}}p_{\vec{l},\vec{k}}X^{\vec{l}}X^{\vec{p}}Z^{\vec{k}}Z^{\vec{q}}(X^{\vec{a}}Z^{\vec{b}})Z^{\vec{q}}Z^{\vec{k}}X^{\vec{p}}X^{\vec{l}}\nonumber\\
&=\sum_{\vec{l},\vec{k}}p_{\vec{l},\vec{k}}X^{\vec{l}+\vec{p}}Z^{\vec{k}+\vec{q}}(X^{\vec{a}}Z^{\vec{b}})Z^{\vec{q}+\vec{k}}X^{\vec{p}+\vec{l}}\nonumber\\
&=\Sigma_k\circ\mathcal{P}_{U}^{(i)}(X^{\vec{a}}Z^{\vec{b}})\label{eq:channel-commutation}\,.
\end{align}
\normalsize
From~\ref{eq:channel-commutation} we can commute the $\sigma$-Pulses trough the noisy channel so that 
\footnotesize
\begin{align}
    \widehat{\VC_B}=&\PC_U^{(L+2)}\circ\UC_B\circ\PC_U^{(L+1)}\circ\VC_L\circ\PC_U^{(L)}\cdots\circ\PC_U^{(1)}\circ\Sigma_1\circ\VC_1\circ\PC_U^{(0)}\nonumber\\
    =&\PC_U^{(L+2)}\circ\UC_B\circ\Sigma_L\circ\PC_U^{(L+1)}\circ\widetilde{\VC}_L\circ\PC_U^{(L)}\cdots\circ\PC_U^{(1)}\circ\widetilde{\VC}_1\circ\PC_U^{(0)}\nonumber\\
    =&\PC_U^{(L+2)}\circ\widetilde{\UC}_B\circ\PC_U^{(L+1)}\circ\widetilde{\VC}_L\circ\PC_U^{(L)}\cdots\circ\PC_U^{(1)}\circ\widetilde{\VC}_1\circ\PC_U^{(0)}\label{eq:proof-theorem-1}\,.
\end{align}
\normalsize
Equation~\eqref{eq:proof-theorem-1} proves Theorem~\ref{theo:1} as it shows that the noisy channels that implement $V(\thv,\gamv)$ and $V(\widetilde{\thv},\widetilde{\gamv})$ are the same.

\end{proof}

\subsection{Proof of Theorem~\ref{theo:2}}

We now prove Theorem~\ref{theo:2}. Here we show that there exist noisy channels $\NC$ (which include as special case then non-unital Pauli noise channels $\PC_{NU}$) that do not commute with the channels $\Sigma$ that implement the $\sigma$-Pulses and which hence break the parameter symmetries.

\begin{proof}
Let us consider a general noisy channel which acts on an $n$-qubit bitstring as
\begin{align}
\mathcal{N}(X^{\vec{a}}Z^{\vec{b}})&=\sum_{\vec{k}, \vec{l}}c_{\vec{k},\vec{l}}^{\vec{a},\vec{b}}X^{\vec{k}}Z^{\vec{l}}\,,\label{eq:nonPauli}
\end{align}
which has some $c_{\vec{k},\vec{l}}^{\vec{a},\vec{b}} \neq 0$ for $\vec{k}, \vec{l} \neq \vec{a}, \vec{b}$. A non-unital Pauli channel in particular must have $c^{\vec{0}, \vec{0}}_{\vec{k}, \vec{l}} \neq 0$ for some $\vec{k}, \vec{l} \neq \vec{0}, \vec{0}$ (see Eq.~\ref{def:non-un-Puali2} of the main text). Similarly to the previous section, we now can define noisy version of the channel that implements $V_B(\thv,\gamv)$  as 
\begin{equation}
    \widehat{\VC_B}=\NC^{(L+2)}\circ\UC_B\circ\NC^{(L+1)}\circ\VC_L\circ\NC^{(L)}\cdots\circ\NC^{(1)}\circ\VC_1\circ\NC^{(0)}\,,
\end{equation}

Let us now analyze if the channel $\Sigma_k$ commutes with a noise channel $\NC$ when acting on an $n$-qubit bitstring $X^{\vec{a}}Z^{\vec{b}}$. 

\small
\begin{align}
    &[\Sigma_k, \NC] (X^{\vec{a}}Z^{\vec{b}}) = (\Sigma_k \circ \NC -\NC \circ \Sigma_k) (X^{\vec{a}}Z^{\vec{b}}) \nonumber\\
    &=  (-1)^{\vec{a}\cdot\vec{q}}(-1)^{\vec{b}\cdot\vec{p}} \sum_{\vec{k}, \vec{l}} c^{\vec{a},\vec{b}}_{\vec{k}, \vec{l}} X^{\vec{k}}Z^{\vec{l}} - 
     \sum_{\vec{k}, \vec{l}} c^{\vec{a},\vec{b}}_{\vec{k}, \vec{l}} (-1)^{\vec{k}\cdot\vec{q}}(-1)^{\vec{l}\cdot\vec{p}} X^{\vec{k}}Z^{\vec{l}} \nonumber\\
    &=  \sum_{\vec{k}, \vec{l}} \left [(-1)^{\vec{a}\cdot\vec{q}}(-1)^{\vec{b}\cdot\vec{p}} c^{\vec{a},\vec{b}}_{\vec{k}, \vec{l}} - (-1)^{\vec{k}\cdot\vec{q}}(-1)^{\vec{l}\cdot\vec{p}} c^{\vec{a},\vec{b}}_{\vec{k}, \vec{l}} \right ] X^{\vec{k}}Z^{\vec{l}}\nonumber \\
    &=  \sum_{\vec{k}, \vec{l}} \left [(-1)^{\vec{a}\cdot\vec{q}+\vec{b}\cdot\vec{p}} - (-1)^{\vec{k}\cdot\vec{q}+\vec{l}\cdot\vec{p}} \right ] c^{\vec{a},\vec{b}}_{\vec{k}, \vec{l}} X^{\vec{k}}Z^{\vec{l}}\nonumber \\
    &= \sum_{\vec{k}, \vec{l}} M^{\vec{a},\vec{b}, \vec{k}, \vec{l}}_{\vec{p}, \vec{q}} c^{\vec{a},\vec{b}}_{\vec{k}, \vec{l}} X^{\vec{k}}Z^{\vec{l}}
\end{align}
\normalsize
where
\footnotesize
\begin{equation}
    M^{\vec{a},\vec{b}, \vec{k}, \vec{l}}_{\vec{p}, \vec{q}} 
    = \begin{cases}
        2 & \text{if}\; \vec{a}\cdot\vec{q} = \vec{b}\cdot\vec{p} \; (\text{mod}\, 2)\, \text{and}\, \vec{k}\cdot\vec{q} \neq \vec{l}\cdot\vec{p} \; (\text{mod}\, 2),\\
        -2 & \text{if}\; \vec{a}\cdot\vec{q} \neq \vec{b}\cdot\vec{p} \; (\text{mod}\, 2)\, \text{and}\, \vec{k}\cdot\vec{q} = \vec{l}\cdot\vec{p}\; (\text{mod}\, 2),\\
        0 & \text{otherwise}.\\
    \end{cases}
\end{equation}
\normalsize
Thus, given $\vec{p}, \vec{q}$ and $\vec{a},\vec{b}$, there exists noise channel defined by some $c^{\vec{a},\vec{b}}_{\vec{k}, \vec{l}}$ such that the $\sigma$-Pulse does not commute with the noise channel. That any such noise channel must be non-Pauli is clear since $M^{\vec{a},\vec{b}, \vec{k}, \vec{l}}_{\vec{p}, \vec{q}} = 0$ for all $\vec{p}, \vec{q}$ whenever $\vec{a},\vec{b} = \vec{k}, \vec{l}$.

\end{proof}

\end{document}